\documentclass[prf,superscriptaddress,showpacs,byrevtex]{revtex4}

\usepackage{graphicx} 
\usepackage{dcolumn}
\usepackage{bm}
\usepackage{color}
\usepackage{float}
\usepackage{hyperref}
\hypersetup{
    colorlinks,%
    citecolor=blue,%
    filecolor=blue,%
    linkcolor=blue,%
    urlcolor=blue
}
 
\usepackage{subfig}

\begin{document}
\title{ Instabilities of MHD flows driven by traveling magnetic fields}

\author{Sandeep R. Kanuganti}
\affiliation{Laboratoire de Physique Statistique, Ecole Normale Superieure, CNRS UMR 8550, 24 rue lhomond 75005
Paris, France
}

\author{Stephan Fauve}
\affiliation{Laboratoire de Physique Statistique, Ecole Normale Superieure, CNRS UMR 8550, 24 rue lhomond 75005
Paris, France
}

\author{Christophe Gissinger }
\affiliation{Laboratoire de Physique Statistique, Ecole Normale Superieure, CNRS UMR 8550, 24 rue lhomond 75005
Paris, France
}

\date{\today}

\begin{abstract}


The flow of an electrically conducting fluid driven by a traveling magnetic field imposed at the endcaps of a cylindrical annulus is numerically studied. At sufficiently large magnetic Reynolds number, the system undergoes a transition from synchronism with the traveling field to a stalled flow, similar to the one observed in electromagnetic pumps. A new type of boundary layer is identified for such electromagnetically-driven flows, that can be understood as a combination of Hartmann and Shercliff layers generated by the spatio-temporal variations of the magnetic field imposed at the boundaries. An energy budget calculation shows that energy dissipation mostly occurs within these boundary layers and we observe that the ohmic dissipation $D_\eta$ always overcomes the viscous dissipation $D_\nu$, suggesting the existence of an upper bound for the efficiency of electromagnetic pumps. Finally, we show that the destabilization of the flow occurs when both dissipations are nearly equal, $D_\nu\sim D_\eta$.
\end{abstract}

\maketitle

\section{Introduction}\label{sec:int}

Magnetohydrodynamics (MHD), i.e. the study of the flow of electrically conducting fluids, play a major role in several natural systems or industrial setups~\cite{Davidson2001}, from the dynamics of flux expulsion involved in solar flares, to the electromagnetic casting of liquid metals in metallurgy~\cite{Asai2000}. It is also an interesting example of energy conversion: In the presence of an electrically conducting fluid, it is possible to generate magnetic energy from the kinetic energy of the fluid, or alternatively, to put in motion a liquid metal by the action of the Lorentz force.  In nature, such a transformation is almost always associated with an instability of the system. The best illustration is dynamo action, i.e. the self-generation of a magnetic field by the action of a turbulent conducting fluid in planetary cores~\cite{Moffatt1978}. In this case, the conversion of fluid motion into magnetic energy is only observed above a given threshold of the fluid velocity~\cite{dynamos}. 

But MHD instabilities can also play the opposite role, by inhibiting or limiting the conversion process, as observed  in electromagnetic linear induction pumps (EMPs). Such EMPs, sometimes called Einstein-Szilard pumps~\cite{Einstein1930}, generally involve liquid metal confined in a cylindrical annular channel subject to a sinusoidal magnetic field imposed by external coils. As the applied traveling magnetic field (TMF) travels with wave-speed $c$  in the axial direction, the liquid metal is pumped at a velocity  $U<c$. The use of such EMPs is of great interest in several industrial processes. In metallurgy for instance, EMPs are widely used to pump molten aluminum during processing of various metals. Another example is the use of EMPs to pump liquid sodium in the secondary cooling circuit of fast breeder reactors, mainly because of the ease of maintenance compared to mechanical pumps \cite{Kliman79}. In all these cases, one objective is to generate a highly efficient pumping of the liquid metal over a relatively large range of operating parameters such as the size of the pump or the developed flow rate.

Interestingly, the efficiency of electromagnetic pumps exhibits a strong limitation when induction effects become dominant over the dissipation of the magnetic field. In large scale pumps, an instability is produced in the flow, which dramatically reduces the developed flow rate: the fluid suddenly bifurcates from a state of quasi-synchronism with the traveling magnetic field toward a strongly stalled regime where the fluid is almost at rest. Gailitis and Lielausis  \cite{Gailitis76} first  formulated a theoretical explanation for this instability. By ignoring the boundary layers generated close to the walls and assuming a velocity field invariant in the radial direction, they were able to propose a criterion for the occurrence of the pump instability. According to this model, the destabilization of the pump should take the form of an inhomogeneity of the flow in the azimuthal direction when the magnetic Reynolds number $Rm_s$ becomes large enough. Here, $Rm_s=(c-U)L/\eta$ is the magnetic Reynolds number based on the slip velocity $c-U$, and $L$ and $\eta$ are respectively a typical scale and the magnetic diffusivity of the fluid.  Recently, it was shown through numerical simulations \cite{Rodriguez2016} that annular EMPs indeed becomes unstable through this scenario, but exhibit an inhomogeneity of the flow in the radial direction, rather than the azimutal one predicted by \cite{Gailitis76}, at least close to the onset. Additional insights on the linear stability of these flows have also be obtained theoretically by investigating more precisely the radial structure of the velocity field, but only in the limit of small magnetic Reynolds number \cite{Zikanov2007,Chu98}.

Experimental studies \cite{Kirillov80} have shown that  low-frequency pulsation in the pressure and the flow rate is indeed observed when induction effects are important, strongly reducing the efficiency of the pump. If the magnetic field from the coils or the sodium inlet velocity are non-uniform in the azimuthal direction, this pulsation takes the form of non-axisymmetric vortices in the annular gap of the pump \cite{Araseki04}. On the other hand, the efficiency of EMPs is also reduced at low magnetic Reynolds number by a strong pulsation at double supply frequency (DSF)~\cite{Araseki00}. 

Despite these theoretical and experimental studies, the exact mechanisms by which the stalling instability occurs at large $Rm$ in the presence of turbulence are still unclear and represent an active field of research. In particular, it was recently shown \cite{Gissinger2016} that this behavior can be connected to flux expulsion: at large magnetic Reynolds number, the driving magnetic field can be suddenly expelled from the conducting fluid, similarly to the skin effect produced by an alternating electric current flowing in a conductor. The corresponding Lorentz force in the bulk flow strongly decreases, and can no longer drive the fluid. The basic aspects of flux expulsion in conducting fluids are relatively well known \cite{Zeldovich1957, Parker1963, Proctor1979}. However, the occurrence of flux expulsion through an instability of the flow remains poorly understood. Because of this crucial role of flux expulsion, the problem of electromagnetic pumps is very close to a theoretical problem studied by Kamkar and Moffatt \cite{Moffatt82}, in which  the flow of a conducting fluid is confined in a rectangular channel and subject to both a transverse pressure gradient in the longitudinal direction and a traveling magnetic field applied on the vertical walls. A transition very similar to the stalling of EMPs was predicted, in which the flow bifurcates from an Hartman-like flow  to a Poiseuille-like flow as the magnetic Reynolds number is increased. Numerical simulations of this problem recently confirmed the role played by flux expulsion in this instability \cite{Bandaru15}. 

In this article, we aim at studying several new aspects of such electromagnetically-driven (EMD) flows. First, the energy budget of these flows exhibits interesting properties: as the energy is injected by a magnetic forcing (through a time-dependent magnetic field imposed at the boundaries), there are different possible routes for dissipating this injected power: it can be directly dissipated through ohmic dissipation without involving the fluid motion, or on the contrary be conversed into kinetic energy through the Lorentz force to be finally dissipated by viscosity. There is currently no general mechanism describing the ratio between viscous and ohmic dissipation in electromagnetically-driven flows. Similarly, what bounds the maximum efficiency of such magnetic to kinetic energy transformation is mostly unknown. Second, the problem involves complex boundary layers due to the inhomogeneous traveling  magnetic field which is applied to the walls, and a full characterization of these layers can certainly help understanding the dynamics of EMD flows. As we will see in this article, there is a deep connection among the structure of these boundary layers, the occurence of the stalling instability and the energy dissipation, leading to interesting predictions for the efficiency of such energy transformations.

In section \ref{num_mod}, we first present the numerical model studied in this paper. Section \ref{sync} reports results obtained in the synchronous regime, whereas section \ref{stal} deals with the stalled regime obtained beyond the instability threshold. Section \ref{energy} discuss some aspects of the energy budget and the efficiency of electromagnetically-driven flows, and our numerical results are understood in the framework of a simple reduced model in section \ref{reduced_model}. Concluding remarks are made in the last section.

\section{Numerical model}\label{num_mod}

\begin{figure}[htbp]
\begin{center}
\subfloat[][]{\label{fig:geomA}\includegraphics[scale=0.3]{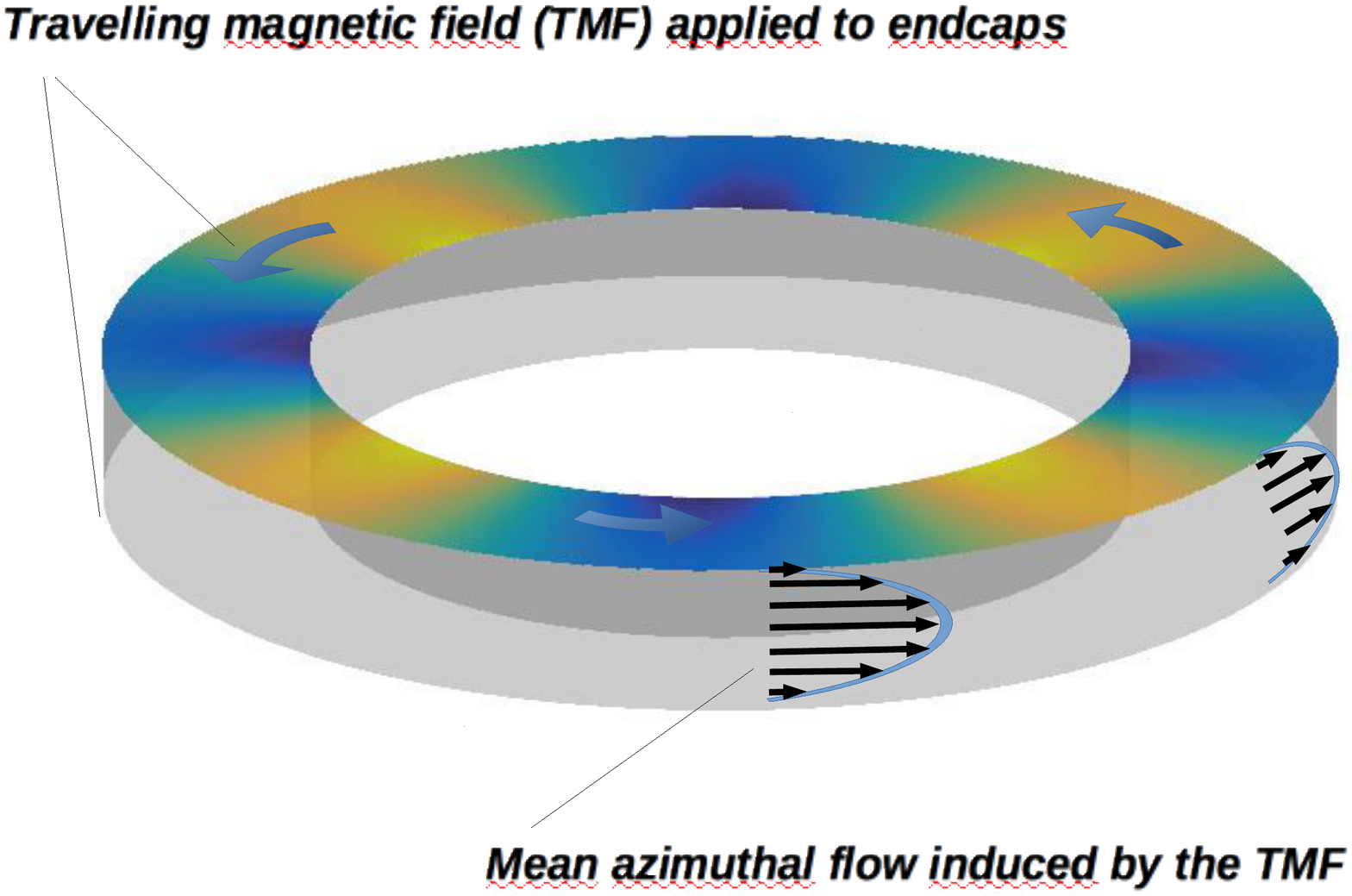}}
\subfloat[][]{\label{fig:geomB}\includegraphics[scale=0.35]{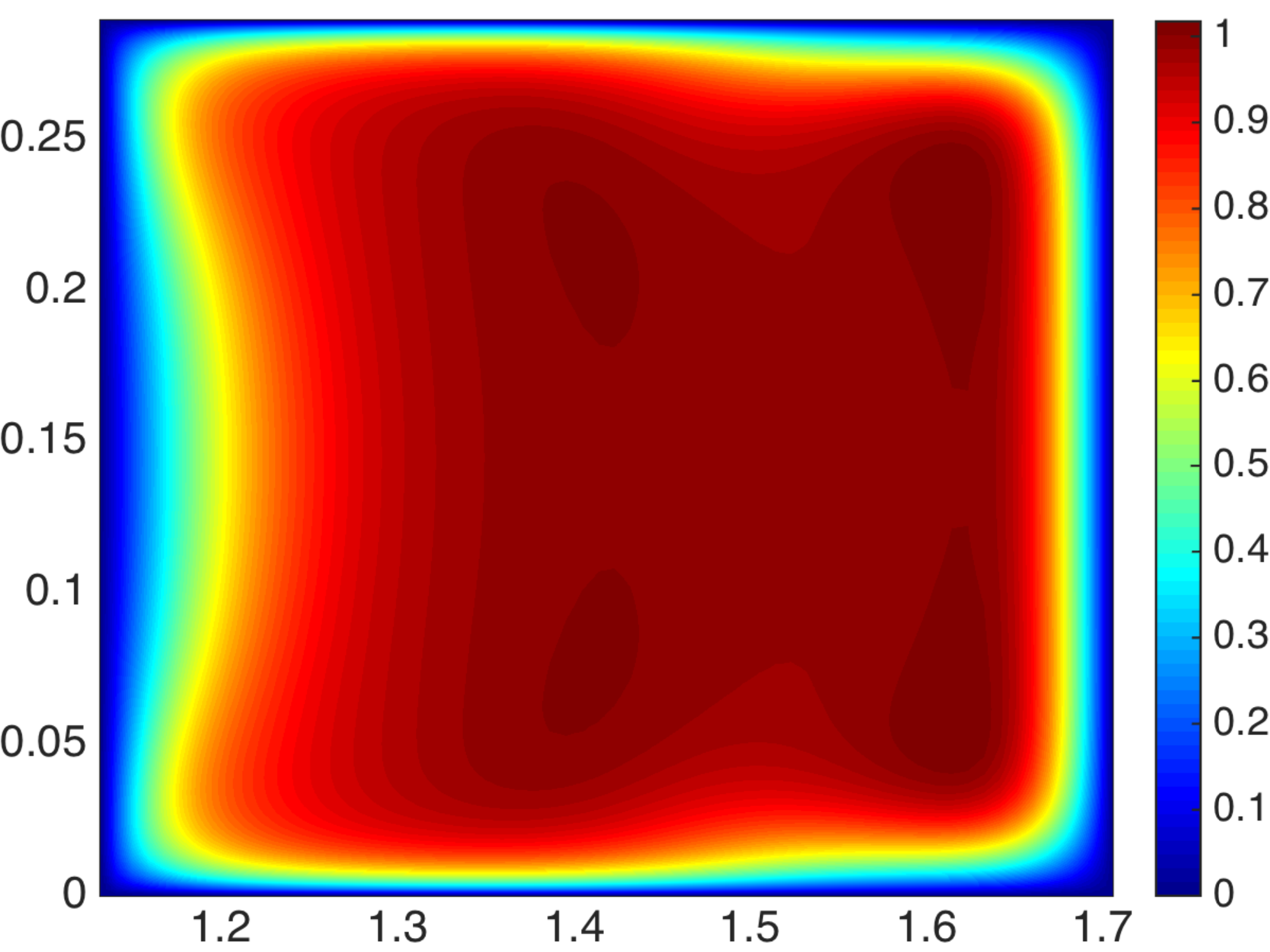}}
\end{center}
\begin{center}
\subfloat[][]{\label{fig:geomC}\includegraphics[scale=0.5]{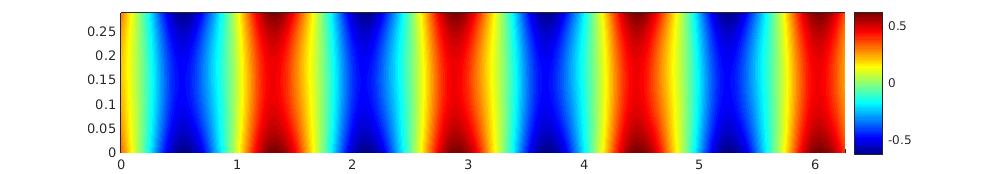}}
\end{center}
\caption{Top-left: geometry of the channel. The fluid is confined between two fixed cylinders and submitted to a sinusoidal magnetic field imposed at the endcaps and traveling in the azimuthal direction. Top-right: azimutal component of the $\theta$-averaged velocity field $U_\theta$ in the $r,z$ plane for $Ha=600$, $Rm=30$ and $Re=4000$. Bottom: corresponding magnitude of the z-component of the magnetic field ($B_z$) in the ($z,\theta$) plane at $r=(r_i+r_o)/2$. A mean flow is therefore generated along the azimuthal direction under the influence of the TMF which is nearly homogeneous in the axial direction.}
\label{fig:geom}
\end{figure}

As shown in the schematic view in Fig.~\ref{fig:geomA}, we consider the flow of an electrically conducting fluid confined between two fixed finite cylinders. $r_i$ is the radius of the inner cylinder, $r_o$ is the radius of the outer cylinder,  $d=r_o - r_i$ is the gap between cylinders, and $H$ is the height of the cylinders. No-slip boundary conditions for the velocity field are imposed at all boundaries.


On the cylinders, we consider high permeability boundary conditions, for which the magnetic field is forced to be normal to the radial boundaries \cite{Gissinger08}. This describes ferromagnetic boundary conditions (sometimes referred as pseudo-vacuum conditions in the literature).  The conducting fluid is only driven by a magnetic forcing, consisting of a radial electrical current $J_s$ imposed on both top and bottom endcaps, such that the boundary conditions on $z=0$ and $z=H$ become:

\begin{eqnarray}
\left({\bf B_2}-{\bf B_1}\right)\cdot{\bf e_z}&=&0\\
{\bf e_{z}}\times \left(\frac{{\bf B_2}}{\mu_2}- \frac{{\bf
    B_1}}{\mu_1}\right)&=&J_s{\bf e_r}
\end{eqnarray}

In the limit of an external ferromagnetic boundary ($\mu_1/\mu_2\ll 1$),
in which the magnetic permeability $\mu_2$ of the boundary is much
larger than the permeability of the fluid $\mu_1=\mu_0$, these conditions reduce to

\begin{eqnarray}
B_\phi(z=0)=\mu_0 J_s  & , &  B_r(z=0)=0,\\
B_\phi(z=H)=-\mu_0 J_s  & , &  B_r(z=H)=0.
\label{BC}
\end{eqnarray}

The electrical surface current $J_s$ is explicitly imposed in the code at both endcaps such that
$J_s=\frac{I_0}{r}\sin\left(m\theta-\omega t\right)$,
where $I_0$ is the amplitude of the applied current, and $\omega$ and
$m$ are respectively the pulsation and the azimuthal wavenumber of
the  traveling magnetic field. Note that due to the solenoidal nature of the magnetic field, these
boundary conditions lead to the generation of a strong axial
magnetic field as well. The flow is therefore driven by the induction due to a sinusoidal magnetic field imposed at the boundaries and traveling in the azimuthal direction.

The governing equations are the magnetohydrodynamic equations, i.e. the Navier-Stokes equations coupled to the induction equation for the magnetic field and the constraint of magnetic field being solenoidal. The system of governing equations are 

\begin{eqnarray}
\rho\frac{\partial {\bf u}}{\partial t} + \rho{\bf u \cdot \nabla u}  &=& -\nabla p + \rho \nu \nabla^2 {\bf u} + {\bf j \times B},  \label{eq:v} \\
\frac{\partial {\bf B}}{\partial t} &=& \nabla \times ({\bf u \times B}) + \frac{1}{\mu_0 \sigma} \nabla^2 {\bf B},  \label{eq:b} \\
\nabla \cdot {\bf B} &=& 0,
\end{eqnarray}
where ${\bf u}$ is the velocity, ${\bf B}$ is the magnetic field, ${\bf j=\frac{1}{\mu_0}\nabla \times B} $ is the electrical current density, $\nu$ is the kinematic viscosity, $\rho$ is the density, $\sigma$ is the electrical conductivity, and $\mu_0$ is the magnetic permeability of vaccum.

In the code, the equations are made dimensionless using a  length scale
$l_0=\sqrt{r_i(r_o-r_i)}$ and a  velocity scale $U_0=c$, where
$c=\omega/k$ is the speed of the  TMF and $k=2m/(r_o+r_i)$. The
pressure scale  is $P_0=\rho c^2$ and the scale of the magnetic field is
$B_0=\sqrt{\mu\rho}c$. The problem is then governed by the geometrical parameters
$\Gamma=H/(r_o-r_i)$ and $\beta=r_i/r_o$ (kept fixed in this paper), the magnetic Reynolds number
$Rm=\mu_0\sigma cl_o$ which compares  induction effects to ohmic dissipation, and the magnetic Prandtl number
$Pm=\nu/\eta$ (ratio of viscosity to magnetic diffusivity $\eta=1/(\mu_0\sigma)$). The magnitude of the imposed current is controlled by
the Hartmann number, defined as
$Ha=\mu_0I_0/\sqrt{\mu_0\rho\nu\eta}$ and comparing the Lorentz force to the viscous force. Alternatively, one may define the kinetic Reynolds number $Re=Rm/Pm$ instead of $Pm$. Note that both $Rm$
and $Re$ are defined here using the speed of the traveling magnetic field and not the one of the fluid. Following \cite{Gailitis76}, we also define the slip magnetic Reynolds number $Rm_s$ defined on the difference of speed between the fluid and the TMF, namely $Rm_s=Rm(1-U/c)$.\\



These equations are integrated with the HERACLES code \cite{Gonzales06}. Originally developed for radiative astrophysical and ideal-MHD flows, it was modified to include viscous and magnetic diffusion, and has been used to describe various problems of diffusive MHD~\cite{Rodriguez2016,Gissinger2016,Gissinger11,Gissinger14}. Note that HERACLES is a compressible code, whereas laboratory experiments generally involve almost incompressible liquid metals. In fact, incompressibility corresponds to an idealization in the limit of infinitely small Mach number ($Ma$). In the simulations
reported here, we used an isothermal equation of state with a small sound speed (in practice $Ma=0.03$), following the approach of \cite{Liu06,Liu08,Gissinger11}. Typical resolutions used in the simulations reported in this article range from $(N_r,N_\theta, N_Z)=128^3$ for laminar flows to $(N_r,N_\theta,N_Z)=[256\times128\times1024]$ for the highest Reynolds numbers. For the velocity field, no-slip conditions are used at both radial and axial boundaries, so the forcing is entirely due to the surface current mentionned above.

%
%

The numerical setup studied here can therefore be regarded as an idealization of electromagnetic pumps. In annular linear EMPs, the flow is confined within a cylindrical channel in which the flow is pumped in the axial direction by a TMF imposed at radial boundaries. In the present setup, the flow is driven in the azimuthal direction, which avoid inlet and outlet effects generated in these industrial pumps. On the other hand, it is also very close to the configuration studied by Kamkar and Moffatt, except that the periodicity of the duct flow is achieved in our simulations by using a cylindrical vessel. Note however that in contrast to \cite{Moffatt82}, no pressure gradient is imposed on the flow, which is solely driven by the TMF.

\section{Synchronous state}\label{sync}

We first describe the basic state obtained at low magnetic Reynolds number and  large Hartmann number, parameters for which the mean flow rate is the largest. In this {\it synchronous} regime, the velocity of the fluid is almost equal to the speed of the traveling magnetic field. In Fig.~\ref{fig:geomC}, we show the magnitude of the z-component of the magnetic field for $Ha=600$, $Rm=30$ and $Re=4000$. Although the applied currents are located at the endcaps boundaries, the magnetic field penetrates throughout the entire section of the channel. Sufficiently far from these boundaries, the field lines become mostly vertical. In particular, field lines are not bent by the flow. The situation is therefore very similar to the one in which a traveling magnetic field $B_z$ would be imposed across the annular channel, which is often considered as an idealization of EMPs. Figs.~\ref{fig:geomB} shows  the azimuthal velocity $U_\theta$ in the $r-z$ plane, averaged in the $\theta$-direction. In the bulk flow, the normalized velocity is very close to $1$, as the flow is almost synchronous with the traveling magnetic field. For some parameters (low $Rm$, very large $Ha$), we have observed that the local velocity can sometimes exhibit values exceeding the speed of the TMF, leading to normalized velocity around $1.4$. This 'superrotation' is very localized in the middle of the channel and strongly depends on the parameters: its study is beyond the scope of the present paper and will be reported elsewhere. 

In addition, very steep velocity gradients are observed near both radial and axial boundaries. The flow profile in the $z$-direction therefore seems very close to a Hartmann-type M-shaped profile. For this reason, such inductive flows are often described in the literature as Hartmann flows, i.e. constant bulk velocity with thin Hartmann boundary layers. Analytical models in which only the time-averaged Lorentz force is taken into account, indeed predict a boundary layer thickness scaling as $1/Ha$, in agreement with Hartmann's prediction. Yet, there is no obvious reason for such an agreement, since the geometry of the traveling magnetic field close to the boundary is considerably different from the normal field lines considered in the classical Hartmann problem.

\begin{figure}[H]
 \begin{center}
\includegraphics[scale=0.65]{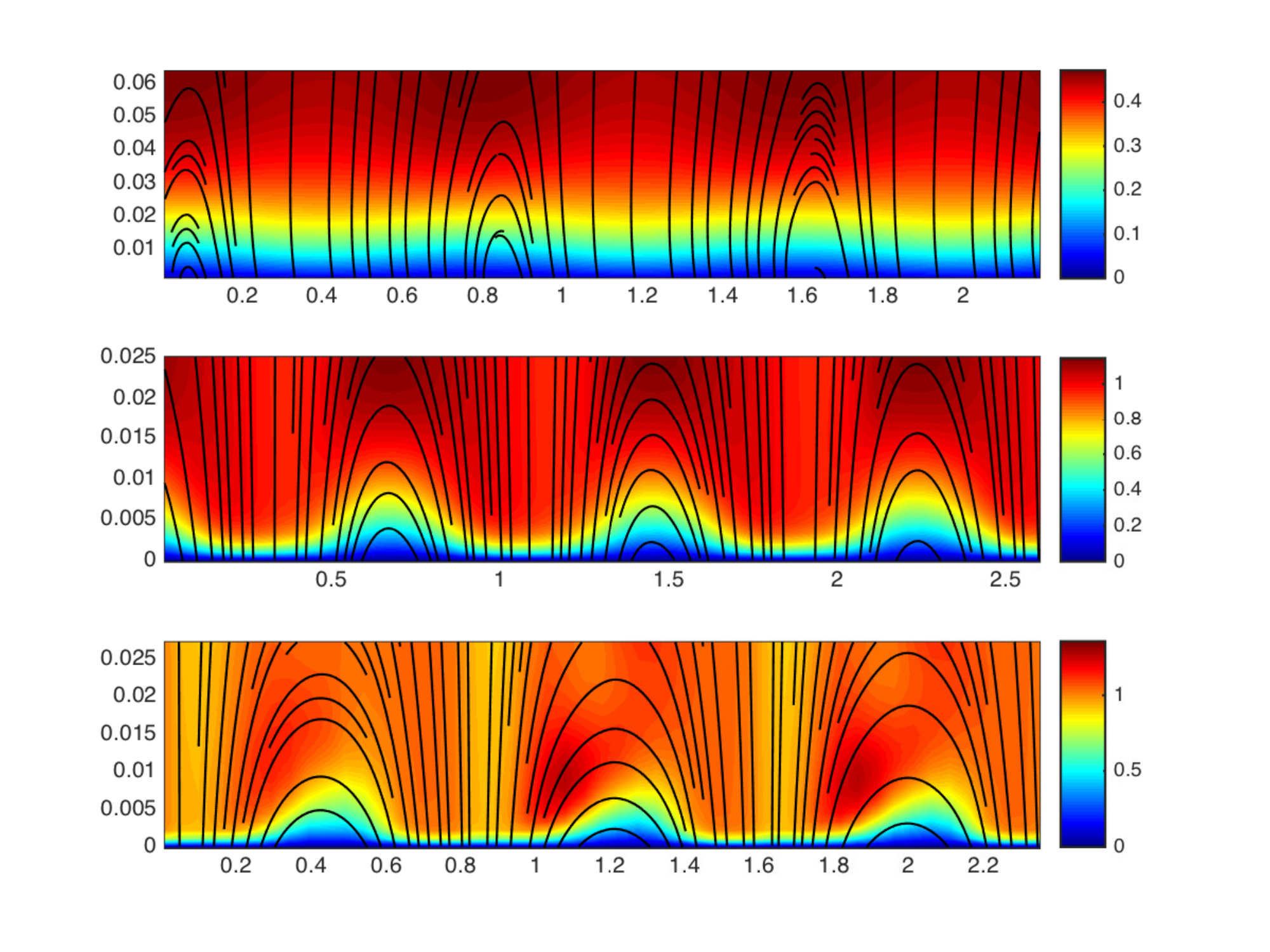}
 \end{center}
 \vspace{-1 cm}
\caption{Structure of the boundary layer close to the bottom boundary, in the synchronous regime, for $Re=4000$, $Rm_s=15$ and three different values of the Hartmann number :$Ha=100$ (top), $Ha=2000$ (middle) and $Ha=10^4$ (bottom). The figure shows colorplots of the velocity component $U_\theta$  and magnetic field lines in the $(z,\theta)$ plane. Note how the local thickness depends on the geometry of the local magnetic field.}  
 \label{fig:pcolor_bls}
 \end{figure}

To investigate further this aspect, we therefore analyzed in more details the structure of the boundary layers generated in our simulations. In Fig.\ref{fig:pcolor_bls}, colorplots of the velocity field close to the lower boundary are shown, together with the field lines of the traveling magnetic field at a given time, for different values of the Hartmann number. Contrary to the well-known Hartmann problem, the thickness of the boundary layer exhibits strong variations  along the azimutal direction. For low magnitudes of the field, these variations are relatively small, but tend to increase at larger $Ha$. Note also that from a sinusoidal variation at low $Ha$, the thickness evolves towards a much more complicated structure for strong fields: regions in which the magnetic field lines are normal to the boundary systematically correspond to a thin boundary layer, while thicker layers are obtained where the field lines are tangential to the boundary.

\begin{figure}[H]
 \begin{center}
\subfloat[][]{\label{fig:layersA}\includegraphics[scale=0.4]{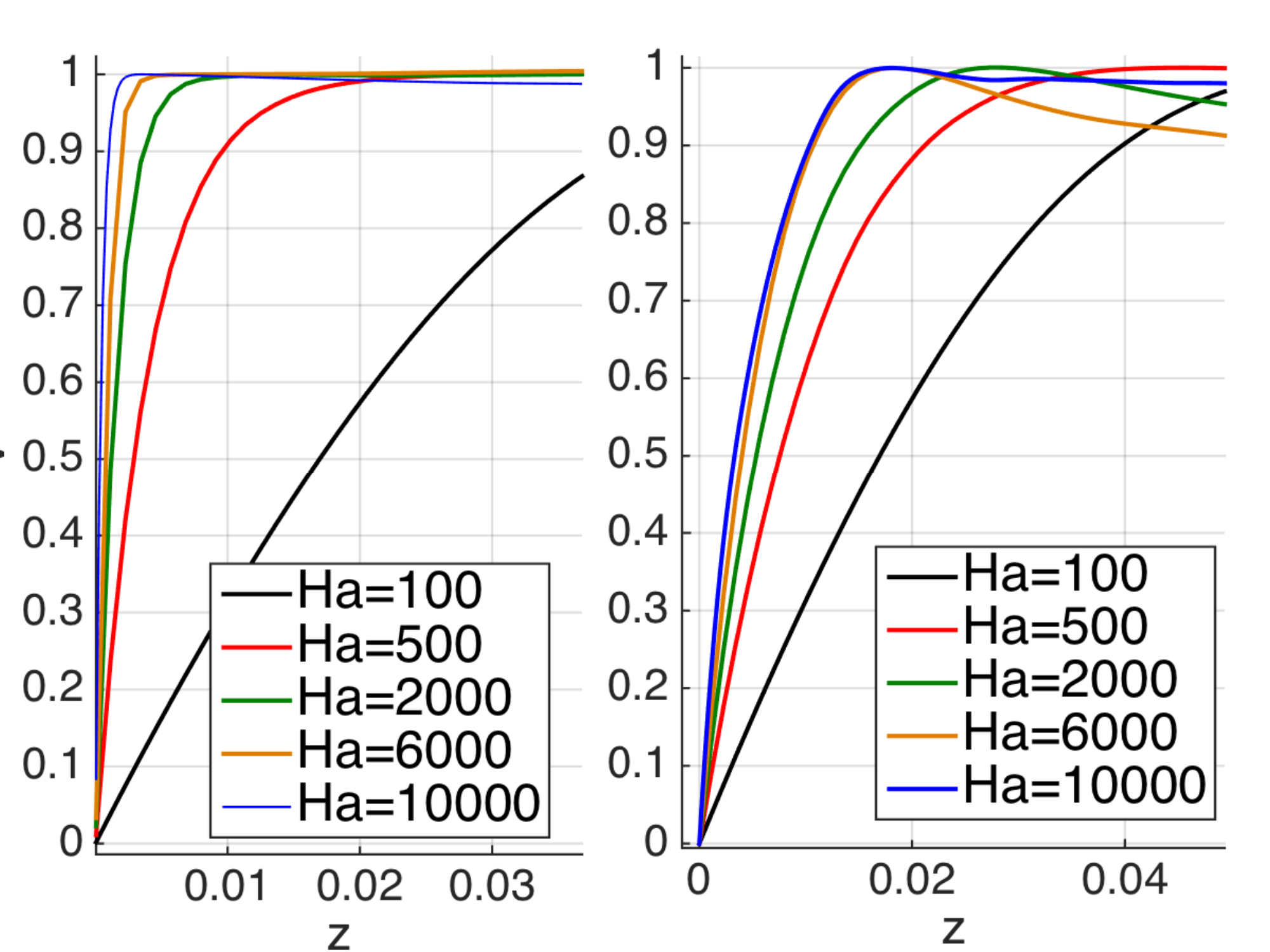} }\\
\subfloat[][]{\label{fig:layersB}\includegraphics[scale=0.4]{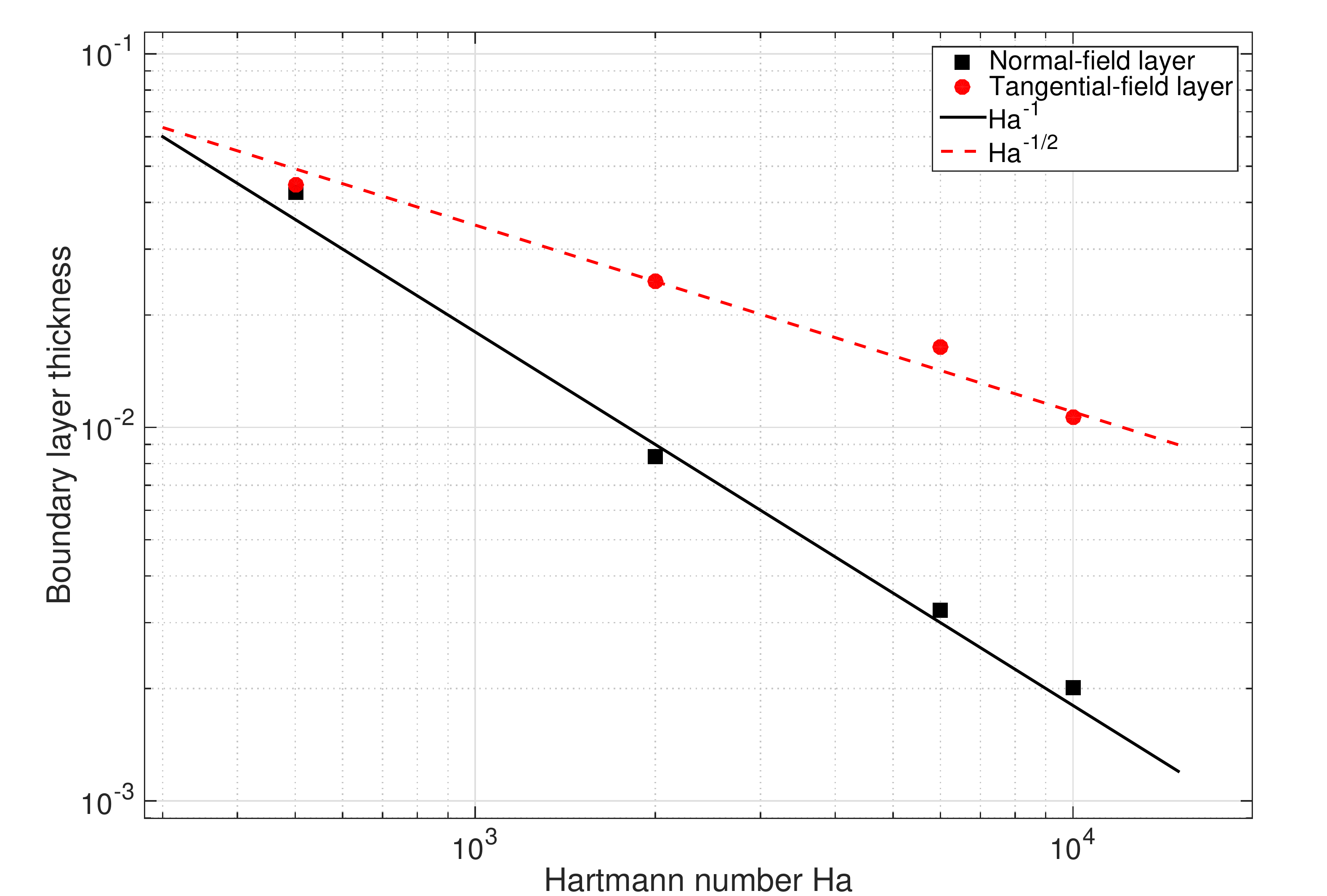}}
 \end{center}
\caption{ (a): Axial evolution of the azimuthal velocity for $\theta$ corresponding  to normal field (left) or tangential field (right), for $Re=4000$, $Rm_s=15$ and various values of the Hartmann number (in the synchronous regime). (b):~Evolution of the boundary layer thickness $\delta$ as a function of the Hartmann number in both cases. Solid black line indicates the Hartmann scaling $\delta\sim1/Ha$, while the dashed red line is the Shercliff scaling $\delta\sim 1/\sqrt{Ha}$. Note the two different behavior, illustrating the existence of a complex 'Schercliff-Hartmann' boundary layer. }  
 \label{fig:layers}
 \end{figure}

 To quantify the evolution of the layer thickness, we plot in Fig.\ref{fig:layersA}  the axial profile of the azimuthal velocity (normalized by its maximum value immediately outside the boundary layer) for both situations, i.e. we show velocity profiles as a function of $z$ for a given value of $\theta$, corresponding either to thin (left) or thick (right) boundary layer regions, for $Re=4000$, $Rm_s=15$ and various values of $Ha$. Each profile is relatively close to an exponential behavior sufficiently close to the boundary. To evaluate the thickness of the boundary layer, we measure the distance across the layer from the boundary to the point where the flow has reached $0.99u_0$, where $u_0$ is the maximum 'free stream' velocity reached outside the boundary layer.
 
Interestingly, the two regions exhibit very distinct scaling laws with the Hartman number. In  regions in which the magnetic field is tangential to the boundary, the boundary layer thickness $\delta$ approximately scales as $1/\sqrt{Ha}$, corresponding to Schercliff's prediction for a shear flow subject to an homogeneous magnetic field parallel to the wall~\cite{Shercliff1953}. On the contrary, a much steeper evolution of the flow is observed in thin regions where the magnetic field is normal to the wall, namely $\delta\sim 1/Ha$,  corresponding to the Hartmann scaling~\cite{Hartmann1937}. 

This boundary layer is very similar to the Roberts layer which arises when a horizontal pipe flow is subjected to a vertical homogeneous magnetic field~\cite{Roberts1967}. The scaling of the layer is however different. In Roberts layers, the thickness of singularities occurring where the field is tangential to the wall scales as $Ha^{-2/3}$ over a distance along the wall scaling as $Ha^{-1/3}$. None of these characteristics are observed in our case, as shown in Fig.\ref{fig:layersB} by the clear-cut scaling $Ha^{-1/2}$. There might be several reasons for such a disagreement: the geometry is relatively different, and the complex structure of the field along the wall is quite different from the simple sinusoidal variation of the angle between the field lines and the wall in Roberts layers. In presence of a traveling magnetic field, the alternation of normal and tangential field close to the boundary therefore  induces a periodic variation of the boundary layer, which oscillates between Hartmann-like and Shercliff-like behavior.


This peculiar geometry of the boundary layer have important consequences on the dynamics of the flow.
For instance, the viscous and ohmic dissipations are predominantly produced in the thinner regions of the boundary layer, where the vorticity and the electrical currents are mainly generated. As we will see in the last section, an Hartmann scaling for the typical dissipation length therefore needs to be invoked to understand the energy dissipation in such electromagnetically-driven flows. Note also that with the flow being mainly driven by induction, the magnetic Reynolds number may also influence the thickness of the boundary layer. As long as the flow is synchronous with the traveling field, we have measured almost no evolution of the boundary layer thickness with $Rm$. As shown in the next section, the situation becomes different as the system deviates from synchronism and expulsion of the magnetic field from the channel becomes significant.

\section{Stalled regime}\label{stal}

As discussed in the introduction, an instability of the flow is expected in EMPs when the magnetic Reynolds number $Rm_s$ becomes large enough: the synchronous regime described above looses its stability and bifurcates towards a non-magnetized state, characterized  by a much smaller flow rate.  A similar transition has been described theoretically in an idealized cartesian channel, in which flux expulsion was identified as the triggering mechanism \cite{Moffatt82}. This instability also occurs in the annular geometry studied here, as illustrated by Fig.\ref{fig:flowrate} showing the evolution of the total azimuthal flow rate across the channel as a function of $Rm$, for different values of $Ha$, $Re$ and the azimuthal wave number $m$ of the applied current. At low $Rm$ and large $Ha$, the velocity is nearly synchronous with the TMF. As  $Rm$ is increased, the synchronism is slowly lost if $Ha$ is not too large, and the flow rate gradually decreases to very small values. For the largest values of $Ha$, this loss of synchronism takes the form of a sharp and hysteretic bifurcation. The mechanism by which such a bifurcation occurs is similar to the one responsible for the staling of annular electromagnetic pumps \cite{Gissinger2016}: on the synchronous branch, the system lies on a stable fixed point, in which the strong Lorentz force balances the viscous dissipation, such that a decrease of the flow speed is always associated to an increase of the accelerating Lorentz force due to the magnetic tension. Above a critical $Rm_s$, any perturbation in which the bulk flow slows down implies an increase of the slip, and therefore leads to a stronger shear of the magnetic field lines.  As a consequence, the magnetic field is expelled away from the bulk flow and the total accelerating Lorentz force decreases, which in turns enhances the initial braking of the fluid, leading to the stalling instability observed in Fig.\ref{fig:flowrate}. 

\begin{figure}[H]
\begin{center}
\includegraphics[scale=0.6]{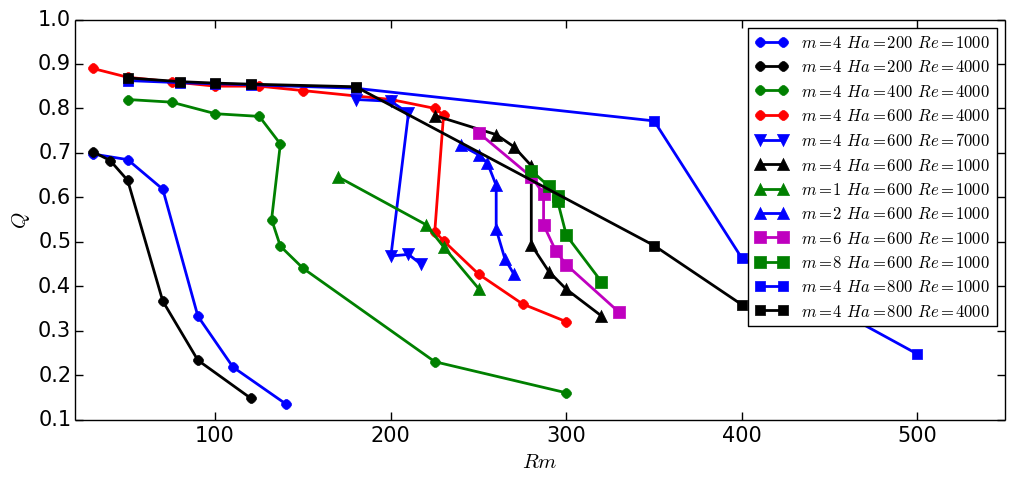}\\
\end{center}
\caption{Total flow rate developed in the channel as a function of $Rm$, for different values of $Ha$, $Re$ and the wavenumber $m$. Note the transition from synchronous ($Q> 0.5$) to stalled ($Q<0.5$) flows as $Rm$ is increased.}  
\label{fig:flowrate}
\end{figure}

\begin{figure}[htbp]
\begin{center}
\subfloat[][]{\label{fig:stalledA}\includegraphics[scale=0.45]{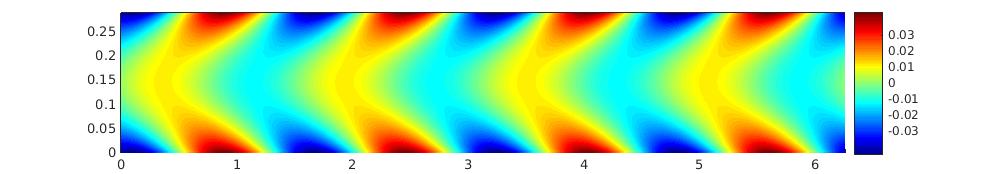}}\\
\subfloat[][]{\label{fig:stalledB}\includegraphics[scale=0.45]{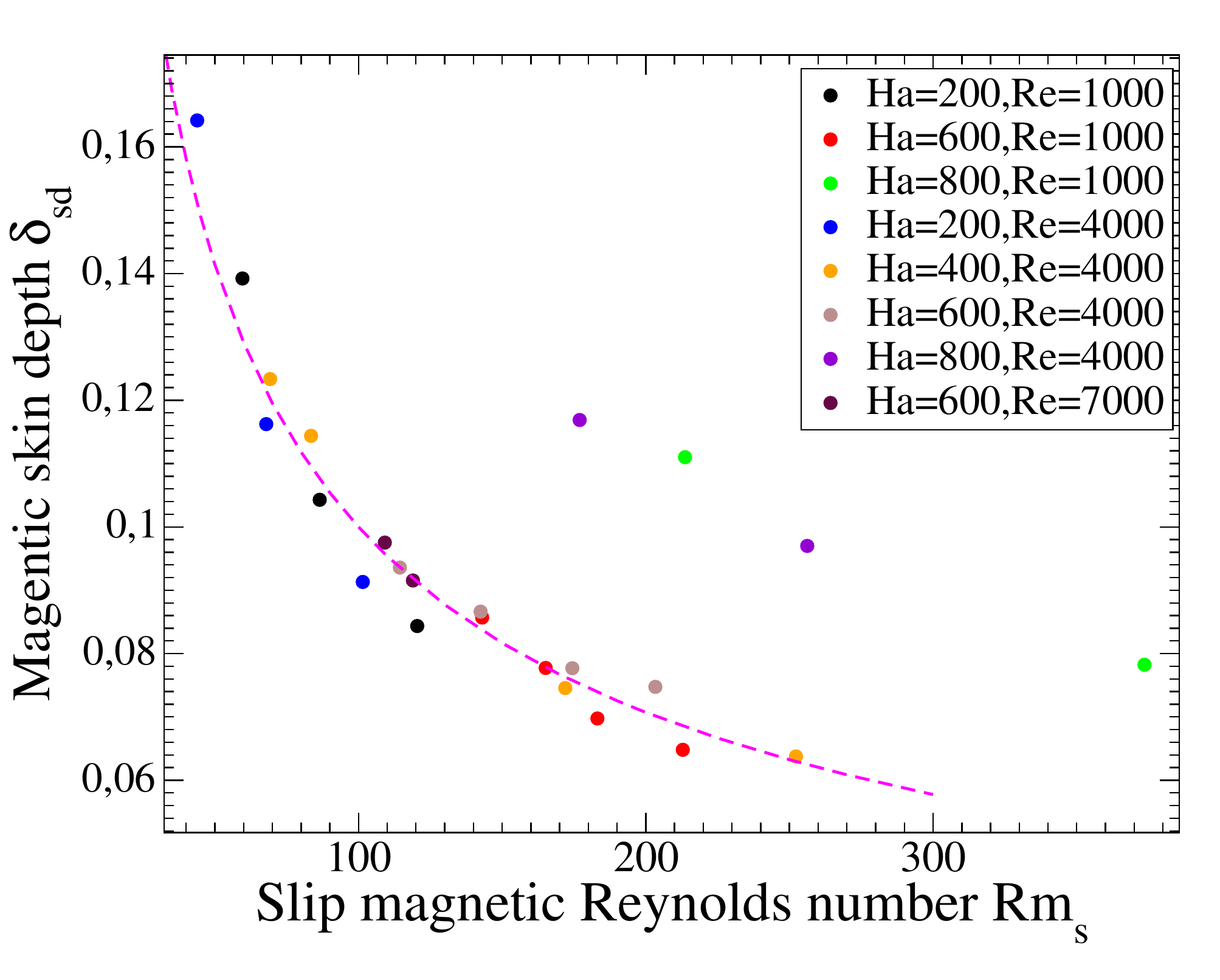}}
\end{center}
\caption{Structure of the magnetic field for $Ha=600$, $Rm=300$ and $Re=4000$. Top : magnitude of the z-component of the magnetic field ($B_z$) in the ($z,\theta$) plane at $r=(r_i+r_o)/2$ .Bottom: Evolution of the skin depth of the magnetic field measured in the simulations as a function of the slip magnetic Reynolds number, for various values of $Ha$ and $Re$ }  
\label{fig:stalled}
\end{figure}

Fig.\ref{fig:stalledA} shows the structure of the magnetic field after the stalling transition, for $Ha=600$, Rm=$300$ and $Re=4000$. The magnitude of the z-component of the magnetic field $B_z$ in the ($z-\theta$) plane is displayed. As expected from the scenario described above, this new state is associated with a magnetic field strongly expelled from the bulk flow and concentrated near the endcaps. In addition, the magnetic field lines are now clearly bent by the difference of velocity between the fluid and the applied traveling field.

The expulsion of the magnetic field away from the bulk can be quantified by measuring the skin depth of the magnetic field after the transition. Fig.\ref{fig:stalledB} shows the evolution of this penetration length $\delta_{B}$ as a function of the slip magnetic Reynolds number $Rm_s$. Although there is still an effect of the Hartmann number, most of the data collapse on the scaling $\delta_{B}\sim 1/\sqrt{Rm_s}$. This scaling law is exactly what is expected for the skin depth generated in a static solid conductor submitted to a rotating magnetic field. This illustrates the fact that flux expulsion is occurring mainly in the bulk flow of the channel.  Note that for the largest value of the Hartmann number ($Ha=800$), both parameters $Ha$ and $Rm_s$ influence the skin depth, leading to a much complex evolution of the system which would certainly need further investigations.


\section{Energy budget}\label{energy}

As shown in the previous section, the problem of electromagnetically driven flows is quite different from Hartmann's solution: induction processes play a major role, and  the power is injected into the induction equation through a magnetic forcing, rather than a body force externally applied to the flow. In the previous section, we showed that  complex boundary layers are generated near the wall where the power is injected. It is therefore interesting to investigate how the energy transfers between kinetic and magnetic terms, and how dissipation is handled in such a system, particularly during the flow instability.

Eqs.~\ref{eq:v} - \ref{eq:b} are used to obtain equations for  magnetic  and  kinetic energy:

\begin{eqnarray}
 \frac{dE^{u}}{dt}&=& L-D_\nu, \label{eq:ke}\\ \frac{dE^{b}}{dt}&=&
 -L - D_\eta+P,\label{eq:me}
\end{eqnarray}
where, $ E^u = \frac{\rho}{2} \langle|\mathbf{u}|^2\rangle_V$ and $E^b =
\frac{1}{2}\langle |\mathbf{b}|^2/\mu_0\rangle_{V}$ are respectively the total kinetic
and magnetic energy densities, $L = \langle\mathbf{u\cdot}[(\mathbf{\nabla\times b)\times
    b}]/\mu_0\rangle_V$ is the power of the Lorentz force and $D_\nu =
\rho\nu\langle\mathbf{u\cdot}\triangle\mathbf{u}\rangle_V=\rho\nu\langle\mathbf{(\nabla\times u)}^2\rangle_V$ is the viscous
dissipation. The term $\frac{\eta}{\mu_0}\int_V b.\Delta b=$ splits into two parts: the surface term $P=\frac{\eta}{\mu_0}\langle\mathbf{b\times}\mathbf{(\nabla\times b)}\rangle_S$ is the injected power due to surface currents imposed at endcaps, and the volume term $D_\eta =
\frac{\eta}{\mu_0}\langle (\nabla\times b)^2 \rangle_V$ is the ohmic
dissipation. $\langle\rangle_V$ denotes spatial average over the annulus, while $\langle\rangle_S$ denotes spatial average over the surface of bottom and top endcaps.

Eq.~\ref{eq:me} expresses that the rate of change of the magnetic energy is equal to the difference between the power injected by the surface currents imposed at endcaps and the magnetic energy loss due to ohmic dissipation and the power of the Lorentz force, hereafter called the Lorentz flux. ($L$ has been defined such that $L>0$ corresponds to a transfer of  energy from the magnetic field to the velocity field). Eq.~\ref{eq:ke} shows that the rate of change of the kinetic
energy is the difference between the Lorentz flux and the viscous dissipation. Therefore,  the Lorentz flux acts as a source term for the kinetic energy, in contrast to the classical case corresponding to dynamo action, in which the energy is usually injected in the velocity field (through body forces or moving boundaries), and used to generate a magnetic field.\\

\begin{figure}[H]
\begin{center}
\subfloat[][]{\label{fig:time_budgetA}\includegraphics[scale=0.4]{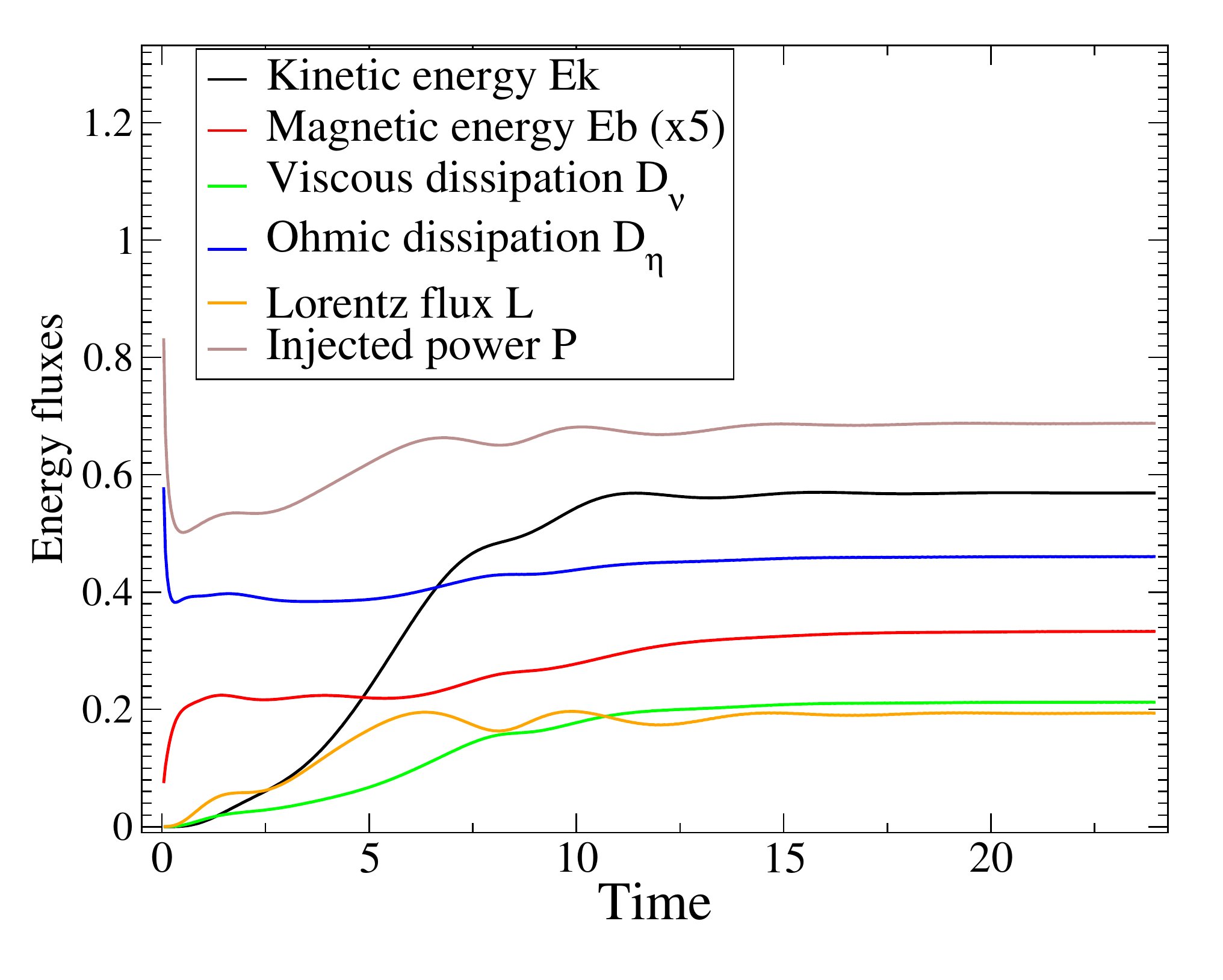}}
\subfloat[][]{\label{fig:time_budgetB}\includegraphics[scale=0.4]{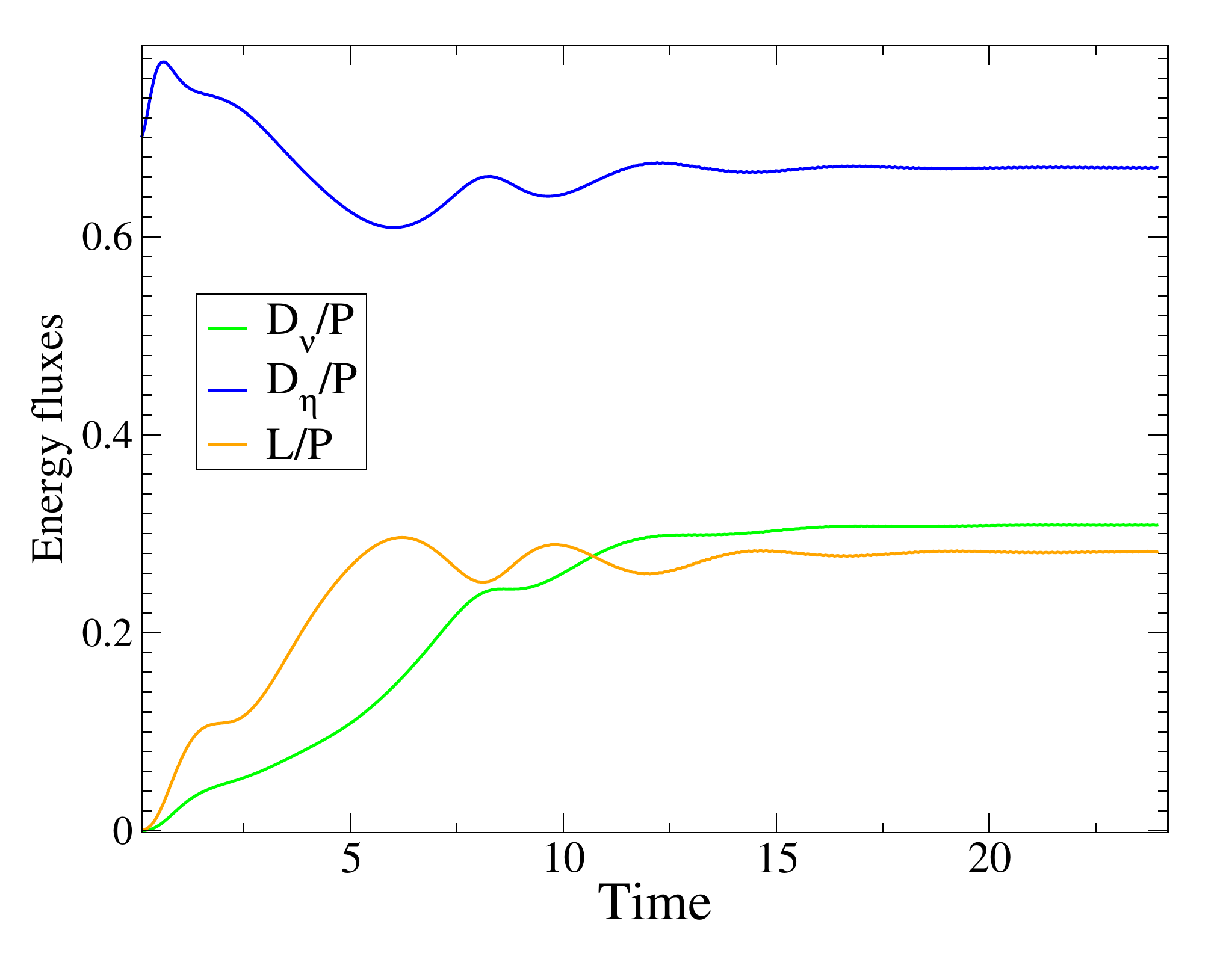}}
\end{center}
\caption{(a): Time evolution of energy fluxes for $Re=4000$, $Ha=600$ and $Rm_s=10$ (synchronous regime). (b):~same thing, but normalized by the injected power.}  
\label{fig:time_budget}
\end{figure}

Fig.~\ref{fig:time_budgetA}  shows the typical time evolution of the energy fluxes for $Re=4000$, $Ha=600$ and $Rm_s=10$, in the synchronous regime. One can see the build-up of kinetic energy as the magnetic forcing establish the mean azimuthal flow in the channel.  An increase of the viscous dissipation is associated with this increase of the kinetic energy. During this phase, both the magnetic energy and the ohmic dissipation stay constant, while the injected power and the Lorentz flux increase as electrical  currents are slowly induced in the channel. Finally,  the saturation is characterized by a small increase of the magnetic energy. When these quantities are normalized by the injected power (Fig.~\ref{fig:time_budgetB}), one can see that the total build-up of the azimuthal flow is characterized by a net transfer from ohmic to viscous dissipation. 

\begin{figure}[H]
\begin{center}
\subfloat[][]{\label{fig:bif_budgetA}\includegraphics[scale=0.35]{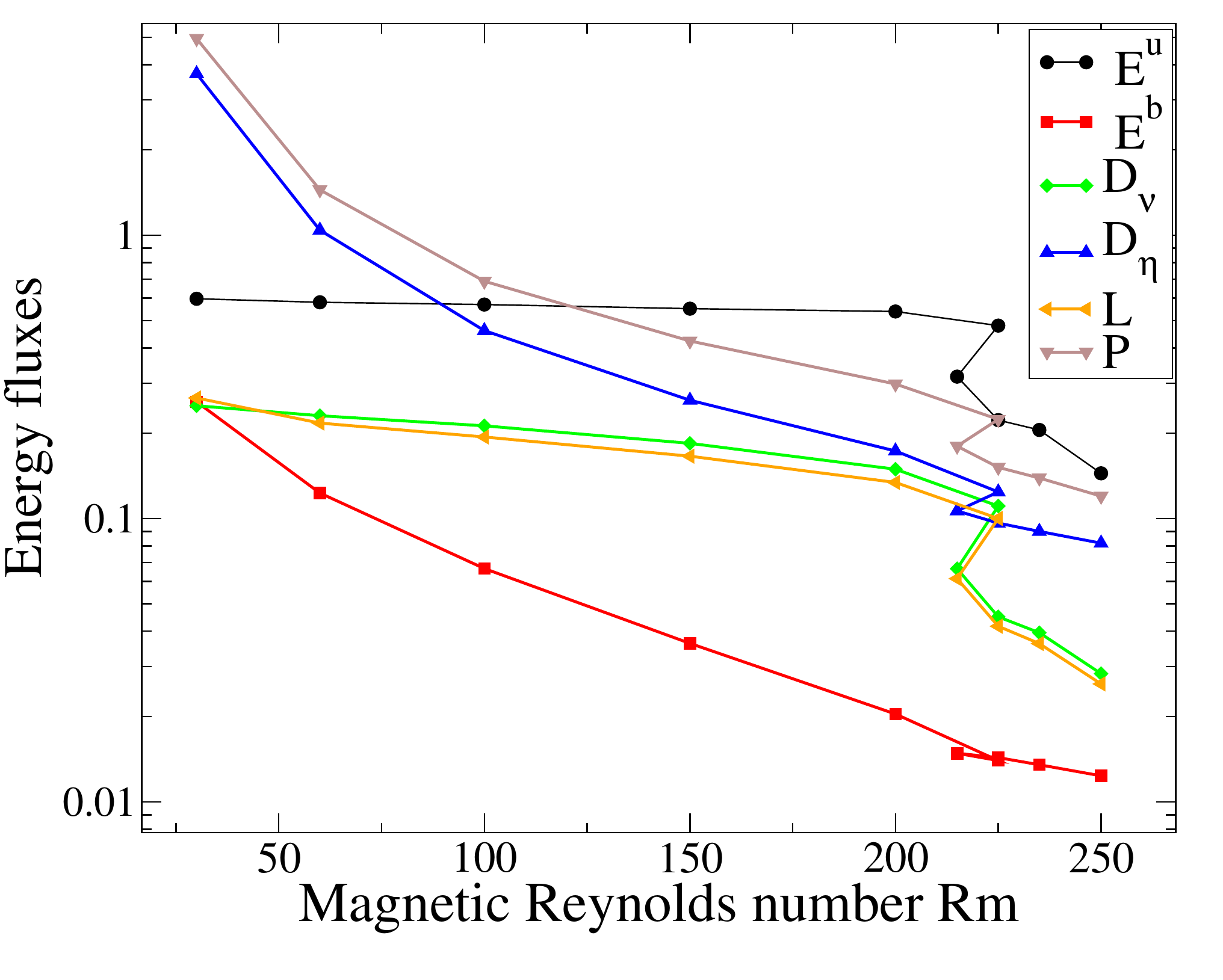}}
\subfloat[][]{\label{fig:bif_budgetB}\includegraphics[scale=0.35]{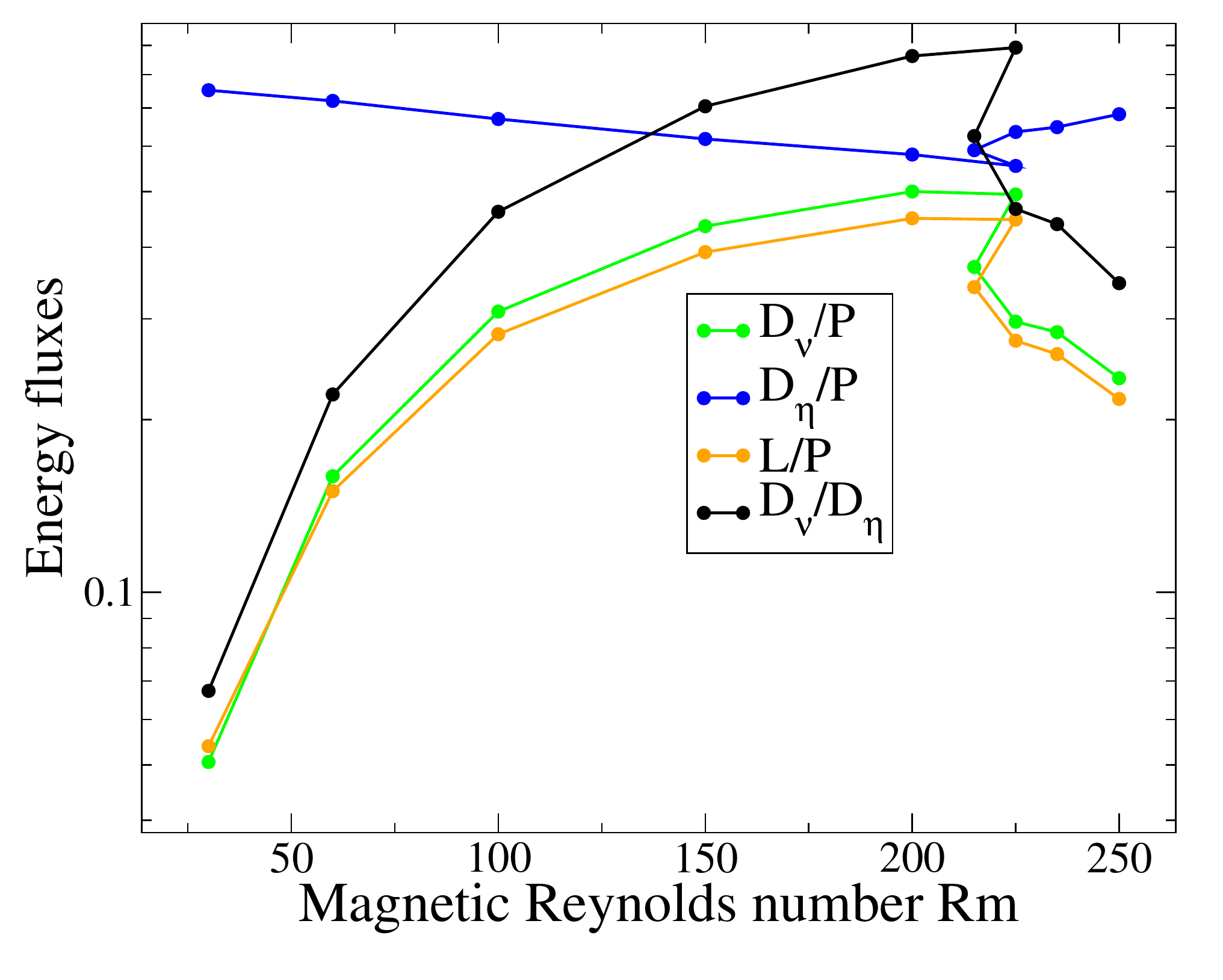}}
\end{center}
\caption{(a): Evolution of energy fluxes for $Re=4000$, $Ha=600$ as a function of $Rm$. (b): same thing, but normalized by the injected power.}  
\label{fig:bif_budget}
\end{figure}

In Fig.~\ref{fig:bif_budget}, we follow the evolution of the saturated values of energy fluxes as a function of $Rm$, for $Re=4000$ and $Ha=600$. The bifurcation from synchronous to stalled flow occurs for $Rm>225$, and is associated with a clear hysteresis. The bistability between the two states is clearly visible on the kinetic energy (black curve), for $Rm=225$.  As the bifurcation point is approached by increasing $Rm$, we observe a decrease of all quantities, including the injected power. At the transition, all the quantities decrease to lower values. Interestingly, the total magnetic energy is associated to a very small hysteresis. It means that during the stalling instability, the magnetic energy is mainly rearranged, all the magnetic field being expelled from the bulk to the boundaries of the channel without changing the total amount of magnetic energy.

For the dissipation, it is more instructive to normalize these quantities by the injected power $P$, as shown in Fig.\ref{fig:bif_budgetB}: as the critical point is approached, the ohmic part of the total dissipation decreases, while the part due to viscous dissipation increases. The normalized Lorentz flux also increases, since it is always balanced with $D_\nu/P$ in steady state. This behavior seems very general: in all the simulations we have performed, the system always dissipates energy preferentially through ohmic dissipation, regardless of the value of $Ha$, $Re$ or even $Pm$, which has been varied between $2.10^{-2}$ and $5$. This is very unusual: in dynamo action for instance, it is known that for a given distance from dynamo onset, this ratio is mainly controlled by the magnetic Prandtl number $Pm$, with dissipation being mainly viscous (resp. ohmic)  at large (resp. small) magnetic Prandtl number $Pm$~\cite{Plunian2010, Brandenburg2014}. \\

\section{Two-dimensional reduced model}\label{reduced_model} 

To understand the existence of a bound on the dissipation ratio, let us write a simple model  in which the cylindrical annulus $(r,\theta,z)$ is approximated by a cartesian channel $(x,y,z)$. Similarly to the above numerical simulations, a surface current ${\bf J_s}=J_0e^{i(ky-\omega t)}{\bf e_x}$  is applied in $z=\pm a$, ensuring the boundary conditions $B_y(\pm a)=\pm B_ae^{i(ky-\omega t)}$ with $B_a$ constant. Given the form of the applied boundary conditions, we seek for a time-harmonic magnetic field of the form ${\bf B}=(B_z(z){\bf e_z}+B_y(z){\bf e_y})e^{i(ky-\omega t)}$, traveling in the $y$-direction. In addition, we will only seek for a purely streamwise time-independent flow such that the velocity depends only on the transverse coordinate, namely ${\bf U}=U(z){\bf e_y}$, and the induced electrical field and  eddy currents are purely spanwise. The calculation is greatly simplified by the use of the vector potential such that  ${\bf B}={\bf \nabla \times}( \tilde{A}(z)e^{i(ky-\omega t)}{\bf e_x})$, and the slip velocity $S=(c-U(z))/c$, where $c=\omega/k$ is the speed of the traveling magnetic field. The induction equation then becomes:

\begin{equation}
\partial_{zz} \tilde{A}=k^2(1-iRS)\tilde{A}\\
\label{induction_model}
\end{equation}
where $R=c\mu_0\sigma/k$ is the magnetic Reynolds number based on the TMF speed. In addition, the electrical currents induced in the channel are given by:

\begin{equation}
\tilde{j}=\frac{1}{\mu_0}\left(\partial_{zz} \tilde{A}-k^2\tilde{A}  \right)=-i\frac{k^2}{\mu_0}RS(z)\tilde{A}(z)
\label{j_model}
\end{equation} 
As we seek for a stationary velocity field, it is necessary to consider only the time-averaged component of the Lorentz force acting on the conducting fluid. With a purely spanwise currents, and looking for a purely streamwise velocity, the mean Lorentz force is therefore expressed as :

\begin{equation}
{\bf F}=-\frac{1}{2}\Re\{{\bf j^*\times} \partial_y{\bf A}\}=\frac{k^3R}{2\mu_0}S(z)|\tilde{A}(z)|^2
\end{equation}
By using this expression of the Lorentz force in the time-averaged Navier-Stokes equation, we finally get:

\begin{equation}
\partial_{zz} S= \frac{k^2}{2\mu_0\rho \nu\eta}S|A|^2
\label{NS_model}
\end{equation}
where $\rho$ and $\nu$ are respectively the density and the kinematic viscosity of the fluid.  Finally,  by using $a$, $c$ and $A_0=B_a/k$ as typical scales for respectively the length, velocity and vector potential, one obtains the following set of equations:

\begin{eqnarray}
\partial_{zz} \tilde{A}&=&\kappa^2\left(1-iRS\right)A \label{mod1},\\
\partial_{zz} S&=& H^2S|A|^2 ,\label{mod2}
\end{eqnarray}
where $H=B_aa\sqrt{\sigma/2\rho\nu}$ is the Hartmann number, and $\kappa=ka$ is the dimensionless wavelength. Equations (\ref{mod1}~-~\ref{mod2}) constitute a closed system of equations for the evolution of both velocity and magnetic field inside the induction channel. If $S$ is supposed constant and equations are integrated along the $z$ direction, this model becomes equivalent to the block velocity model described in \cite{Gailitis76}.\\

{\bf Inductionless limit}: Although this set of equations can be solved numerically, it is instructive to first study the limit $R\ll 1$, in which the induced field stays small compared to the magnetic field  applied at the boundaries. In the long wavelength limit $ka\ll1$, equations (\ref{mod1}, \ref{mod2}) can be solved exactly:

\begin{equation}
U(z)=c\left[1-\frac{\cosh(\frac{H}{\kappa} z)}{\cosh(\frac{H}{\kappa})} \right]  \hspace{2cm}
\end{equation}
The velocity field therefore consists of a bulk flow almost constant and synchronous with the traveling magnetic field, associated to thin Hartmann boundary layers close to the walls. Both viscous and ohmic dissipations can be easily computed from this expression for the velocity field:

\begin{eqnarray}
D_\nu=\rho\nu\int_V({\bf \nabla\times U})^2 dV=\frac{\rho\nu c^2}{a}\left[ H\tanh(H)-\frac{H^2}{\cosh^2(H)} \right]
\label{Du}\\
D_\eta=\mu_0\eta\int_V\left(\frac{{\bf \nabla\times B}}{\mu_0}\right)^2 dV=\frac{\sigma B_0^2c^2}{(ka)^2}\left[ \frac{1}{H}\tanh(H)+\frac{1}{\cosh^2(H)} \right]
\label{Db}
\end{eqnarray}
where the dissipation rates per unit of area have been computed. Hence, the dissipation ratio takes a very simple form:

\begin{equation}
\frac{D\nu}{D_\eta}=\frac{H\tanh(H)-H^2/\cosh^2(H)}{H\tanh(H)+H^2/\cosh^2(H)} < 1
\end{equation}
The above expression shows that, in agreement with the direct numerical simulations, the total dissipation is always dominated by the ohmic losses, and both dissipations asymptotically become equals for sufficiently large Hartmann number. Note that this bound only stands in the limit of weak induction, and concerns the laminar solution. It is deeply linked to the thin boundary layers generated close to the walls, in which most of the dissipation occurs. The first term in the brackets in expressions \ref{Du}-\ref{Db} can be regarded as the contribution to the dissipation from the boundary layers, while the second term represents the energy dissipated by the bulk flow. When the boundary layer becomes sufficiently thin, the energy is almost entirely dissipated within these boundary layers, and both dissipations become equals. The physical reason for such equality  can be understood  by the following heuristic argument: Let us suppose that most of the velocity gradients are confined inside the boundary layer. In this case, all the viscous dissipation is achieved inside this boundary layer, and expression (\ref{Du}) simply reduces to  $D_\nu\sim \rho\nu c^2/\delta$, where $\delta\sim a/H$ is the thickness of the Hartmann boundary layer. As the induced current takes the form $j(z)\sim \sigma B_0(U(z)-c)$, the ohmic dissipation can be written as 

\begin{equation}
D_\eta\sim \sigma B_0^2\int_0^a\left(U-c\right)^2 dz\sim \sigma B_0^2\left[\int_0^\delta\left(U-c\right)^2 dz+\int_\delta^a\left(U-c\right)^2 dz\right]
\end{equation}
 The second term is the contribution from the bulk flow, while the first term is the ohmic dissipation inside the boundary layer, and tends to $\sigma B_0^2c^2\delta$ when the bulk flow is nearly in synchronism with the TMF. In the presence of Hartmann boundary layers $\delta\sim a/H$, this last expression is therefore identical to the viscous dissipation. Note however that such an equality between dissipations in the boundary layer only stands for sufficiently small boundary layer thickness, i.e. for sufficiently large Hartmann number. 
 
In any induction machine, an electrical conductor can never move exactly in synchronism with the traveling magnetic field, because it would imply a frame of reference in which the magnetic field is steady: even far from the boundaries, the velocity $u$ therefore has to be smaller than $c$, meaning that non-negligible electrical currents are always generated in the bulk flow. In other words, there is always an additional ohmic dissipation in the bulk flow, making it slightly larger than the viscous one. As the flow tends to synchronism, this contribution becomes negligible and the dissipation ratio reaches one.\\

\begin{figure}[H]
\begin{center}
\includegraphics[scale=0.35]{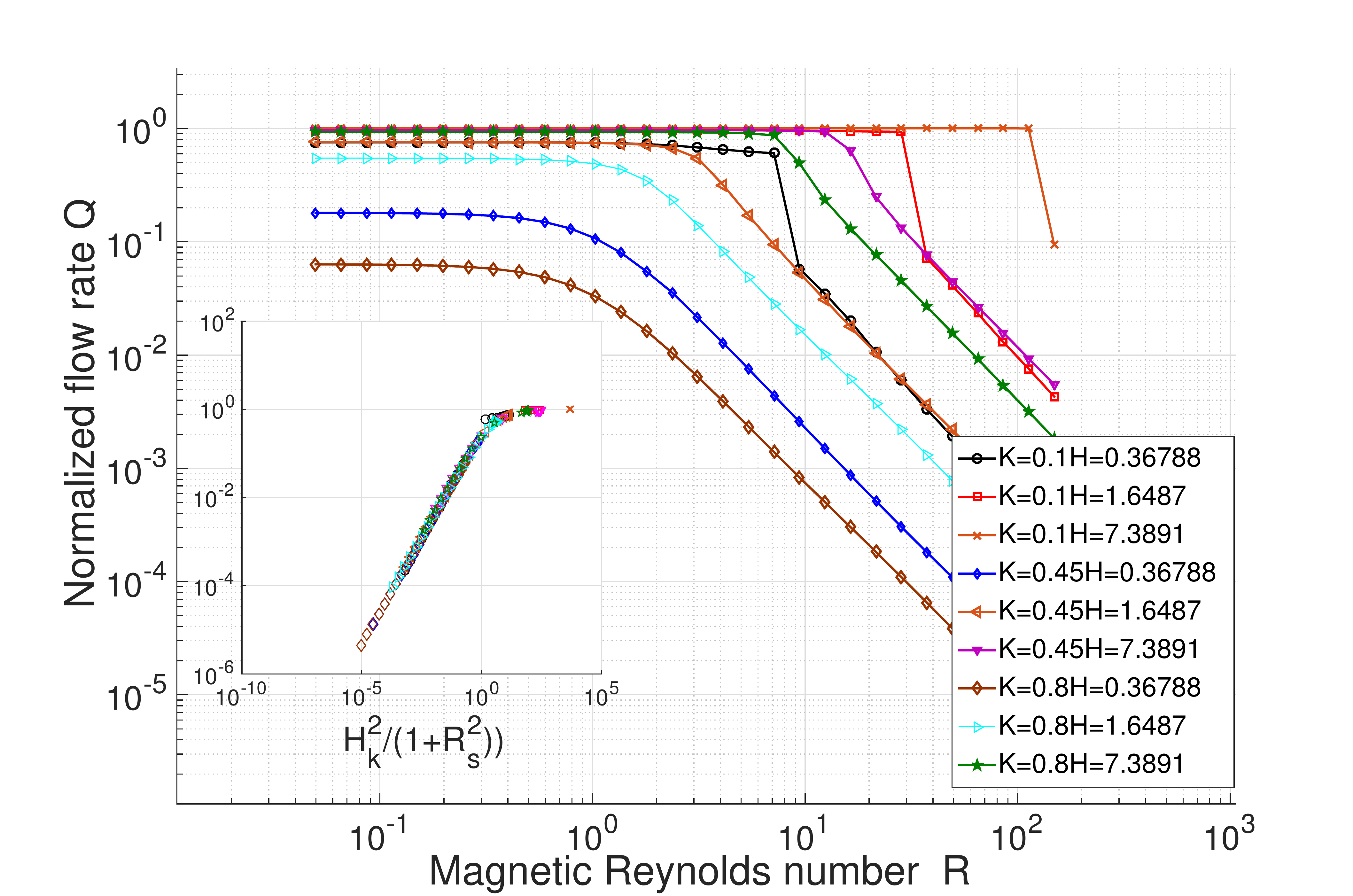}
\end{center}
\caption{Evolution of the dimensionless flow rate $Q$ as a function of $R$, for various values of $H$ and $\kappa$, obtained by numerical integration of the reduced model (eqs (\ref{mod1}-\ref{mod2})). Inset: all the curves collapse when plotted as a function of the dimensionless number $N$ (see text).  }  
\label{fig:model}
\end{figure}

\begin{figure}[H]
\begin{center}
\subfloat[][]{\label{fig:comparaisonA}\includegraphics[scale=0.28]{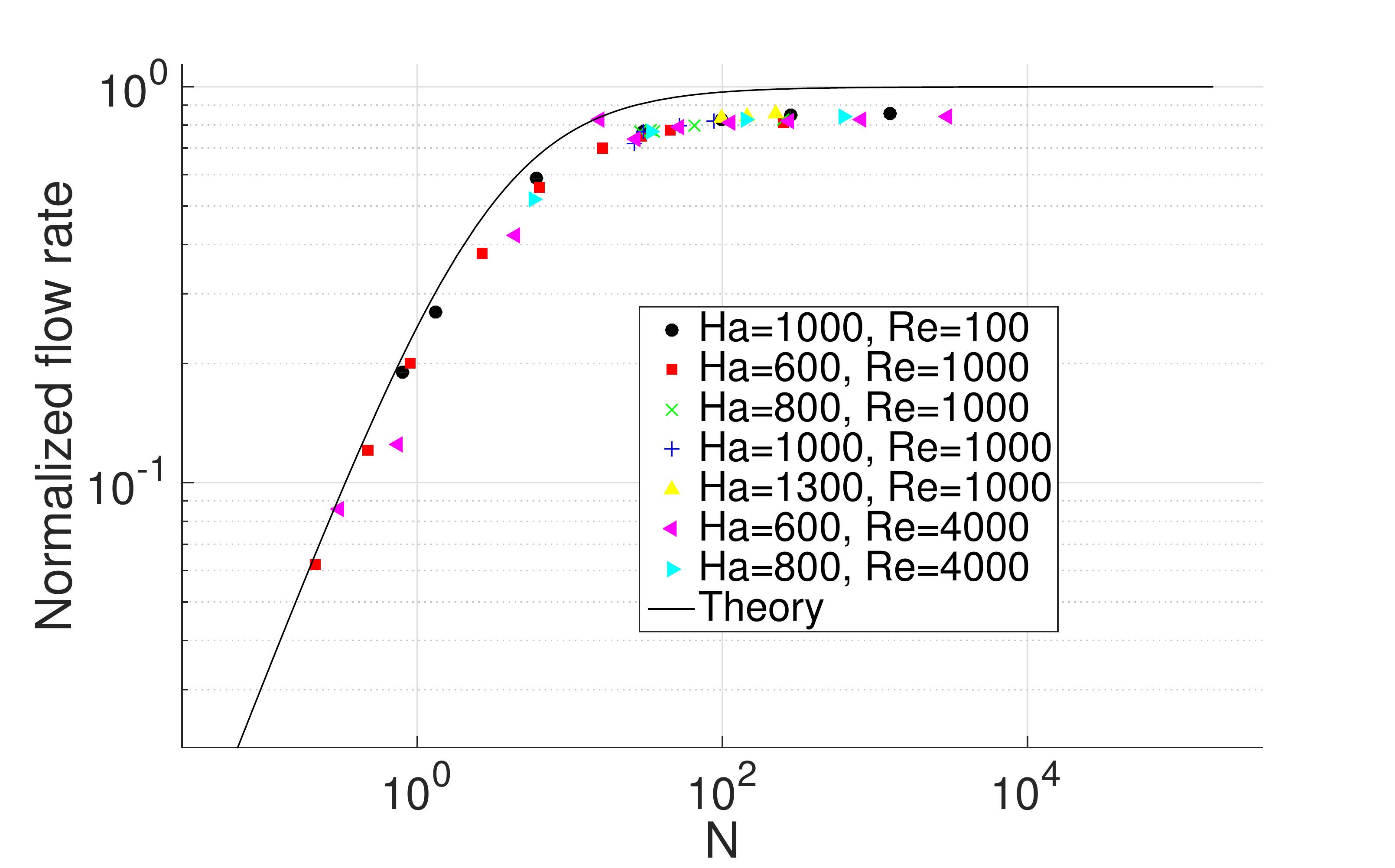}}
\subfloat[][]{\label{fig:comparaisonB}\includegraphics[scale=0.28]{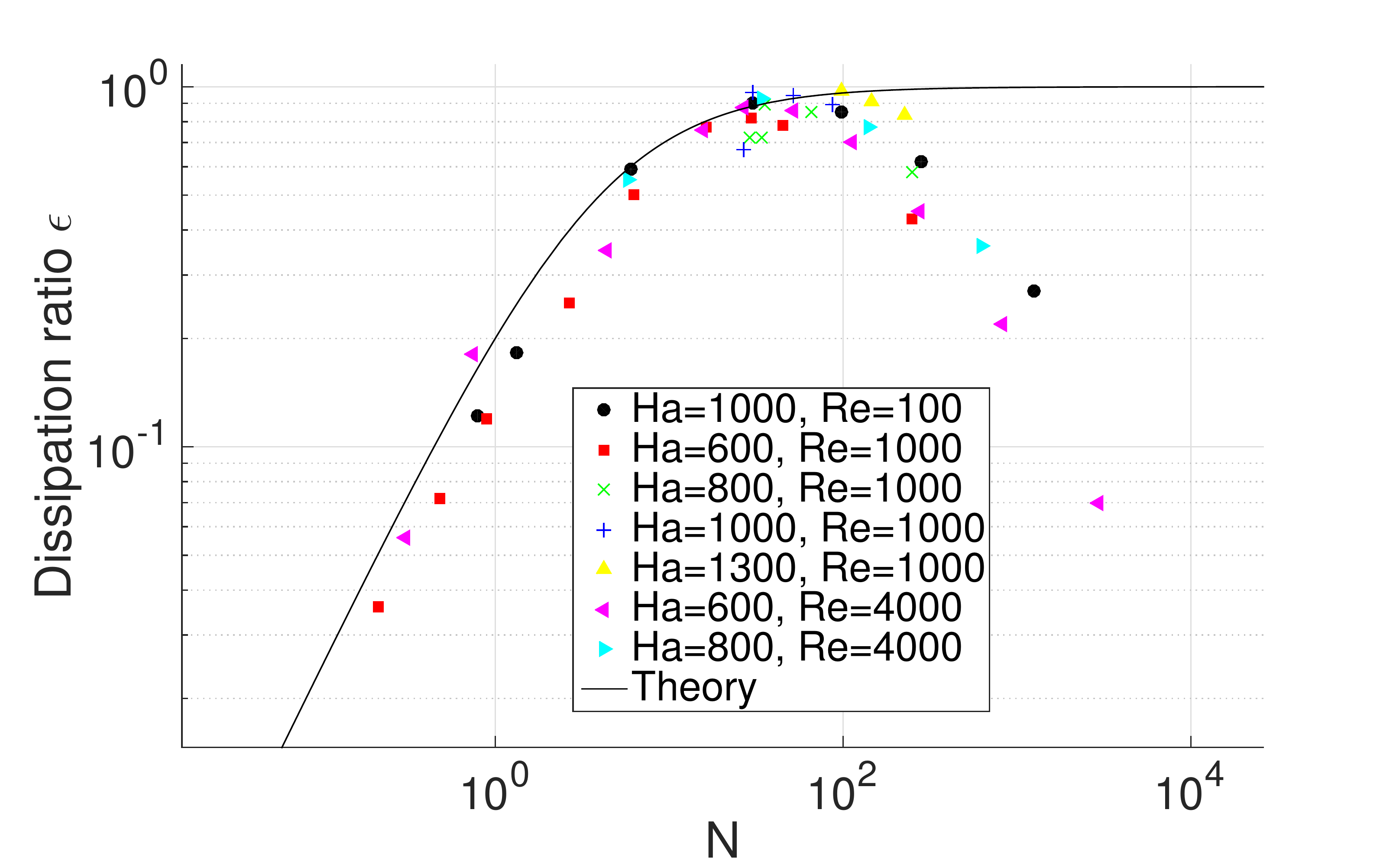}}
\end{center}
\caption{ (a): Total flow rate $Q$ developed in the channel as a function of $N$, for different values of $Ha$, $Re$ and $Pm$. All the numerical points are rescaled on a single curve, corresponding to the reduced model (\ref{mod1}-\ref{mod2}) (black curve). (b): Evolution of the dissipation ratio $\epsilon=D_\nu/D_\eta$ as a function of $N$. Note the bound on the dissipation ratio, maximum for  $N\sim1$. }  
\label{fig:flowrate}
\end{figure}

{\bf General case}: The regime $R\gg 1$ can be investigated by numerical integration of equations (\ref{mod1}~-~\ref{mod2}). Fig.~\ref{fig:model} shows the evolution of the normalized flow rate $Q=\int_{-1}^{1}qdz$ as a function of $R$, for various values of $H$ and $\kappa$. Similarly to the numerical simulations, large Reynolds number and low Hartmann numbers are associated with strongly stalled flows, while the fluid is synchronous with the traveling magnetic field at low $R$ and/or large Hartmann number $H$. This reduced model is very useful to understand the dynamics of the flow: when the model (\ref{mod1}-\ref{mod2}) is integrated along the $z$ direction, and if the velocity slip is supposed constant, the mean force balance between the Lorentz force and the laminar viscous dissipation can be written: 

\begin{equation}
\frac{H^2(1-Q)}{\kappa^2(1+R^2(1-Q)^2)} - Q =0.
\end{equation} 
 where the first term represents the Lorentz force and the second term is the linear viscous dissipation. By defining a magnetic Reynolds number $R_s=(c-u)/(k\eta)$ based on the slip velocity, the evolution of the total flow rate across the channel is therefore given by:
 
 \begin{equation}
 Q=\frac{N}{1+N}  \hspace{1cm} ,  \hspace{1cm} N=\frac{H_k^2}{1+R_s^2}
 \label{modelQ}
 \end{equation}
 where $H_k=B_a\sqrt{\sigma/2\rho\nu}/k$  is the Hartmann number based on the wavelength $k$ of the traveling magnetic field. In the inset, it is shown that all the curves indeed collapse on a single one when $Q$ is  plotted against the dimensionless number $N$. It is now instructive to come back to the direct numerical simulations described in the previous sections. Fig.~\ref{fig:comparaisonA} shows the evolution of the flow rate $Q$ as a function of the dimensionless number $N$. Most of the data are rescaled on a single curve, corresponding to the prediction (\ref{modelQ}). Note that there is no adjustable parameter for the model.  The agreement between the  numerics and our simplified model is very good for small flow rates, but a clear departure from the prediction is observed very close to synchronism, where the flow rate is slightly smaller than predicted. This new parameter $N$ is also useful to understand the bound on the dissipation ratio observed in the simulations. Fig.~\ref{fig:comparaisonB} shows the dissipation ratio $\epsilon=D_\nu/D_\eta$ as a function of  $N$ for both the direct simulations and the reduced model. First, the solid line shows that the bound on the dissipation ratio obtained in the limit $R\ll1$ still holds when equations \ref{mod1}-\ref{mod2} are integrated for finite values of $R$. In addition, although very different values of $Re$ and $Ha$ are used ($Pm$ is varied from $2.10^{-2}$ to $5$.), the dissipation ratio $\epsilon$ is solely controlled by $N$, and systematically smaller than unity, regardless of the control parameters.  
 
Finally, two different regimes can be observed in Fig.\ref{fig:comparaisonB}. For small values of $N$, a very good agreement with our theoretical model is obtained: the dissipation ratio scales as $\epsilon\sim N^2$ at small $N$ and saturates to $1$ at larger $N$. For larger values of $N$ however, there is a strong discrepancy between the theory and the DNS: The dissipation ratio again decreases towards small values. This may be inferred to several characteristics of the flow which are not taken into account by the model. For instance, at small magnetic Reynolds number, such electromagnetically-driven flows exhibit an amplification of the forcing,  known as the double-supply frequency (DSF). This pulsation is known to decrease the efficiency of electromagnetic pumps, and may explain the decrease of $\epsilon$ observed here. Since equations (\ref{mod1}-\ref{mod2})) are obtained by time-averaging the Navier-Stokes equation, this effect can not be described by our reduced model. Note also that the 'Shercliff' part of the boundary layer is not taken into account. Nevertheless, the bound remains identical: the system always dissipates energy preferentially through ohmic dissipation in the thin Hartmann-like boundary layers. Note also that the stalling of the pump seems to always occur in the vicinity of $D_\nu/D_\eta=1$.

\section{Discussion and conclusion}

This work presents a numerical study of the magnetohydrodynamic flow generated  in a cylindrical annulus subject to a traveling magnetic field imposed at the endcaps. We have shown that when the magnetic Reynolds number $Rm_s$ is too large, such electromagnetically-driven flows experience a MHD instability similar to the stalling of electromagnetic pumps, in which the flow rate can suddenly drop to inefficient regimes due the expulsion of the magnetic flux outside the bulk flow. Our analysis highlights the new type of boundary layers generated in such induction-driven flows: due to the spatial inhomogeneity of the applied magnetic field, the boundary layer thickness exhibits azimuthal variations along the wall, alternating thin and thicker layers. Each of these regions scale differently with the Hartmann number depending on the local geometry of the field (normal or parallel to the wall).  This new result has two consequences: Since only the thicker layer contributes to the flow averaged in the azimuthal direction, the boundary layer of the averaged flow exhibits a Schercliff scaling. On the contrary, viscous and ohmic dissipations mainly occur in the thinner part of the boundary layer, where the gradients are the strongest. We have also showed that the penetration of the axial magnetic field into the bulk flow is controlled by the magnetic Reynolds number $Rm_s$ based on the slip velocity between the TMF and the induced flow. The scaling of this skin depth  $\delta_{B}\sim 1/\sqrt{Rm_S}$ is identical to the classical penetration length of a varying magnetic field close to an electrical conductor.

Finally, we performed a study of the energy budget of this electromagnetically-driven flow: we showed that the dissipation ratio $D_\nu/D_\eta$ seems to be  bounded by $1$, meaning that the ohmic dissipation always overcomes the viscous one, even for magnetic Prandtl number larger than unity. This unexpected result is in good agreement with the predictions from a reduced model in two dimensions.

Finally, let us emphasize that this last result has an important consequence on the efficiency of electromagnetic pumps. In steady-state regime, equations (\ref{eq:ke}-\ref{eq:me}) indicate that the efficiency $\gamma=L/P$ of this energy transformation can be related simply to the dissipation: $\gamma=\epsilon/(1+\epsilon)$, where $\epsilon=D_\nu/D_\eta$ is the dissipation ratio. The limitation on $\epsilon$ reported in our numerical simulations and the reduced model suggests the existence of an exact bound for the efficiency of this electromagnetic pump, namely $\gamma<\frac{1}{2}$. Although this bound has been obtained in a very idealized situation, several results reported here should apply beyond our system. For instance, we expect  the mixed Hartmann-Shercliff layer to be generated as soon as an MHD flow is subject to a traveling magnetic field with a sufficient large magnitude. Similarly, in several industrial applications, the magnetic Reynolds numbers $Rm_s$ are of the same order than the ones reported here. One weak point of our numerical analysis is certainly the fluid Reynolds number, many orders of magnitude smaller than in real applications. It would be therefore interesting to see if our efficiency bound still holds for more turbulent flow. In linear induction pumps, the magnetic field is applied only on a section of the channel, such that the efficiency is evaluated by measuring the pressure drop between inlet and outlet. In this perspective, it is interesting to note that an asymptotic efficiency close to $50\%$ seems to be asymptotically reached in several experimental setups  (see for instance Fig.14 in \cite{Fanning2003} or Fig.9 in \cite{Ota2004}), in agreement with our prediction. Such an agreement can be understood with simple arguments. For sufficiently large Reynolds number, both the bulk flow and the boundary layers should be turbulent. In this case, if the same turbulent viscosity is used for describing the bulk flow and the boundary layers, the dissipation ratio should remain bounded by one. It follows that the maximum efficiency of an electromagnetic pump should also be $50\%$ in the fully turbulent regime.
Future work devoted to the optimization of this bound should involve much higher Reynolds numbers in order to describe this regime, and should also include an adverse pressure gradient as in real pumps. It may greatly help improving  the efficiency of many industrial processes.
 \\

{\bf Acknowledgements:}

This work was granted access to the HPC resources of MesoPSL financed  by the Region Ile de France and the project Equip@Meso (reference
ANR-10-EQPX-29-01) of the programme Investissements d'Avenir supervised by the Agence Nationale pour la Recherche. We also benefited from the computational support of the HPC resources of GENCI-TGCC-CURIE (Project No. t20162a7164). Support of the Indo-French Centre for the Promotion of Advanced Research (IFCPAR/CEFIPRA) contract 4904-A is acknowledged.



\begin{thebibliography}
\bibitem


\bibitem{Davidson2001}
P.A. Davidson, {\it An Introduction to Magnetohydrodynamics}, Cambridge: Cambridge University Press (2001)

\bibitem{Asai2000}
S. Asai, {\it Recent development and prospect of electromagnetic processing of materials}, Science and Technology of Advanced Materials, Volume 1, Issue 4, 191-200 (2000)

\bibitem{Moffatt1978}
See, for instance, Moffatt H. K.,{\it Magnetic Field Generation in Electrically Conducting Fluids}, (Cambridge University Press) (1978)

\bibitem{dynamos}
R. Monchaux et al., Phys. Rev. Lett., 98 044502 (2007); R. Stieglitz and U. Muller, Phys. Fluids 13, 561(2001); A. Gailitis et al, Phys. Rev. Lett. 84, 4365 (2000)

\bibitem{Einstein1930}
A. Einstein and L. Szilard, {\it Electrodynamic Movement of Fluid Metals Particularly for Refrigerating Machines,}, Patent Specification 303065, United Kingdom, application date: December 9, 1928, acceptance date: May 26 (1930), 

\bibitem{Kliman79}
G.B. Kliman, {\it Large electromagnetic pump}, Electr. Mach. Electromech., {\bf 3}, 129-€"142 (1979)

\bibitem{Gissinger08}
C. Gissinger, A. Iskakov, S. Fauve, E. Dormy, {\it Effect of magnetic boundary conditions on the dynamo threshold of VonKarman swirling flows}, Euro. Phys. Lett., {\bf 82}, 29001 (2008)

\bibitem{Gailitis76} 
Gailitis A. and O. Lielausis, {\it Instability of homogeneous velocity distribution in an induction-type mhd machine}, Magnetohydrodynamics, {\bf 1},
69-79 (1976). Translation from Magnitnaya Gidrodinamika 1, 87-101 (1975).

\bibitem{Rodriguez2016}
P. Rodriguez-Imazio and C. Gissinger, {\it Instabilities in electromagnetically-driven flows, part II}, Phys. Fluids 28, 034102 (2016)


\bibitem{Zikanov2007}
X. Ai, B.Q. Li, O. Zikanov, {\it Stability of electromagnetically-driven flows in induction channels}, Magnetohydrodynamics  43-1, 63-82 (2007)

\bibitem{Chu98}
S. Cho and S. Hee Hong, {\it the magnetic field and performance calculations for an electromagnetic pump of a liquid metal}, J. Phys. D: Appl. Phys. 31 2754-2759 (1998)

\bibitem{Kirillov80}
Kirillov, I.R., Ogorodnikov, A.P., Ostapenko, V.P. ,{\it Local  characteristics  of  a
cylindrical  induction  pump  for  Rms larger than 1 } ,  English translation from Magnitnaya Gidrodinamika {\bf 2}, 107-€"113 (1980)


\bibitem{Araseki04}
Araseki, H., Kirillov, I.R., Preslitsky, G.V., Ogorodnikov, {\it Magnetohydrodynamic instability in annular linear induction pump:: Part I. Experiment and numerical analysis}, A.P., Nucl. Eng. Des. {\bf 227}, 29-€"50. (2004)

\bibitem{Araseki00} 
Araseki H., Kirillov I., Preslitsky G. and Ogorodnikov A. P., {\it Double-supply-frequency  pressure  pulsation  in  annular
linear  induction  pump.  Part  I.  Measurement  and  numerical
analysis.}, Nucl. Engin. and Design  {\bf 195}, 85-100 (2000)


\bibitem{Gissinger2016}
C. Gissinger, P. Rodriguez-Imazio, S. Fauve,{\it Instabilities in electromagnetically-driven flows, part I}, Phys. Fluids 28, 034101 (2016)


\bibitem{Zeldovich1957}
A.B. Zeldovich RY,  Sou. Phys. J.E.T.P.4, 460-462. (1957) 

\bibitem{Parker1963}
E.N Parker, Astrophys. Journ.138, 552-575 (1963)

\bibitem{Proctor1979}
M. R.E. Proctor and  D.J. Galloway, Journ. Fluids. Mech. 90, 273-287 (1979) 

\bibitem{Moffatt82}
H.Kamkar and H.K.Moffatt, J. Fluid Mech., {\it A dynamic runaway effect associated with flux expulsion in magnetohydrodynamic channel flow}, {\bf 121},107-122 (1982)

\bibitem{Bandaru15}
V. Bandaru, J. Pracht, T. Boeck, J. Schumacher, {\it Simulation of flux expulsion and associated dynamics in a two-dimensional magnetohydrodynamic channel flow}, Theo. Comp. Fluids Dyn. A 29 Issue 4, p263-276. (2015)

\bibitem{Gonzales06}
Gonzales et al,  {\it HERACLES: a three-dimensional radiation hydrodynamics code},  Astronomy and Astrophysics, {\bf 464}, 429-435 (2006).

\bibitem{Gissinger11} 
C. Gissinger, H. Ji and J. Goodman, {\it  The role of boundaries in the Magnetorotational instability}, Phys. Fluids. {\bf 24}, 074109 (2012).

\bibitem{Gissinger14}
C. Gissinger, {\it The Taylor-vortex dynamo}, Phys. Fluids, {\bf 26}, 044101 (2014)

\bibitem{Liu06}
W. Liu, J. Goodman, and H. Ji, {\it Simulations of magnetorotational instability in a magnetized Couette flow}, Astrophys. J. {\bf 643}, 306 (2006).

\bibitem{Liu08}
W. Liu, {\it Numerical study of the magnetorotational instability in Princeton MRI experiment}, Astrophys. J. {\bf 684}, 515 (2008).

\bibitem{Shercliff1953}
Hartmann, K. Dan. Vidensk. Selsk. Mat. Fys. Medd. 15, (1937). 

\bibitem{Hartmann1937}
J.A. Shercliff, Math. Proc. Cambridge Philos. Soc. 49, 136 (1953).

\bibitem{Roberts1967}
P. Roberts, {\it Singularities of Hartmann layers}, Proc. R. Soc. Lond. A, 300 (1967)

\bibitem{Plunian2010}
F. Plunian and R. Stepanov, {\it Cascades and dissipation ratio in rotating magnetohydrodynamic turbulence at low magnetic Prandtl number}, Phys. Rev. E, 82, 4 046311 (2010)

\bibitem{Brandenburg2014}
A. Brandenburg, {\it Magnetic Prandtl number dependence of the kinetic-to-magnetic dissipation ratio}, The Astrop. Mourn., Vol. 791-1 (2014)

\bibitem{Fanning2003}
A. Fanning et al, {\it Giant Electromagnetic pump for sodium cooled reactor applications}, Electric Machines and Drives Conference, 2003

\bibitem{Ota2004}
H. Ota et al., {\it Development of 160 m3/min large capacity sodium-immersed self-cooled electromagnetic pump} , J. Nucl. Sci. Tech., {\bf 41},511-523 (2004)


\end{thebibliography}
\end{document}